\documentclass[amsmath,onecolumn, superscriptaddress,preprint,aps]{revtex4}
\usepackage{amsthm,amsfonts,graphicx,verbatim, color}
\usepackage{graphicx}
\usepackage{bm}
\usepackage[utf8]{inputenc}
\let\oldmarginpar\marginpar
\renewcommand\marginpar[1]{\-\oldmarginpar[\raggedleft\footnotesize #1]%
{\raggedright\footnotesize #1}}

\newcommand{\be}{\begin{equation}}
\newcommand{\ee}{\end{equation}}
\newcommand{\bea}{\begin{eqnarray}}
\newcommand{\eea}{\end{eqnarray}}

\newcommand{\Tr}{{\rm Tr}\,}
\renewcommand{\epsilon}{\varepsilon}

\renewcommand{\cite}[1]{[\onlinecite{#1}]}

\begin{document}

\title{Many-body localization and thermalization in quantum statistical mechanics}
\author{Rahul Nandkishore}
 \affiliation{Princeton Center for Theoretical Science, Princeton University, Princeton, New Jersey 08544, USA}
 \email{rahuln@princeton.edu}
\author{David A. Huse}
 \affiliation{Princeton Center for Theoretical Science, Princeton University, Princeton, New Jersey 08544, USA}
 \affiliation{Department of Physics, Princeton University, Princeton New Jersey 08544, USA}
 \email{huse@princeton.edu}
\begin{abstract}
We review some recent developments in the statistical mechanics of isolated quantum systems.  We provide a brief introduction to quantum thermalization,
paying particular attention to the `Eigenstate Thermalization Hypothesis' (ETH), and the resulting `single-eigenstate statistical mechanics'.
We then focus on a class of systems which fail to quantum thermalize and whose eigenstates violate the ETH:  These are the many-body Anderson localized systems;
their long-time properties are not captured by the conventional ensembles of quantum statistical mechanics.  These systems can locally remember forever information
about their local initial conditions, and are thus of interest for possibilities of storing quantum information.  We discuss key features of many-body
localization (MBL), and review a phenomenology of the MBL phase.  Single-eigenstate statistical mechanics within the MBL phase reveals dynamically-stable
ordered phases, and phase transitions among them, that are invisible to equilibrium statistical mechanics and can occur at high energy and low spatial dimensionality
where equilibrium ordering is forbidden.

\end{abstract}
\maketitle
 \tableofcontents
 \newpage
 \section{Introduction}

Although many of the fundamentals of quantum statistical mechanics were formulated together with those of quantum mechanics itself almost a century ago \cite{sakurai}, the subject has had a recent rebirth.  This has occurred due to the development of methods and tools in atomic, molecular, optical and condensed matter physics that allow one to build, control, and study in the laboratory many new sorts of assemblies of strongly-interacting quantum degrees of freedom \cite{npi}.  Such systems extend our abilities to explore and understand many-body quantum mechanics, and are also of interest due to possibilities of using them as components of new quantum technologies.

  Many formulations of quantum statistical mechanics postulate that the system of interest is in contact with an `external' reservoir \cite{Kardar},
  with certain properties of the reservoir often taken for granted. 
  However, recent experimental progress in well-approximating isolated many-body quantum systems (such as cold gases of neutral atoms \cite{bloch})
  motivates a fresh consideration of the statistical mechanics of {\it closed} quantum systems, i.e. isolated quantum systems not coupled to any external reservoirs.
  [See e.g. \cite{PolkovnikovRMP} for a recent review with some overlap with the present review.]
  The statistical mechanics of closed quantum systems is also important as a point of principle,
  since if we assume that any external reservoir or measuring apparatus is itself a quantum system it can then be
  included as part of the closed many-body quantum system of interest.

  One fundamental question about such a closed quantum many-body system is: What states does its unitary time evolution bring it to after an arbitrarily long time?  To make the possible answers to this question sharp, one needs to consider the thermodynamic limit of a large system, as we will discuss below.  There appear to be two answers to this question that are robust under small but arbitrary local perturbations to the system's Hamiltonian, namely {\it thermalization} and {\it localization}.  The answer can depend on the nature of the system and on the initial state being considered, and the system can show quantum phase transitions between these two possibilities as the system or the initial state is varied.  The main goal of the present paper is to define, discuss and elaborate these two possibilities, reviewing some of what is either known or (mostly) conjectured about them.

  In order for a closed quantum system to thermally equilibrate under its own dynamics, the system must be able to act as its own reservoir, i.e. the
  dynamics must be such that for a subsystem that contains only a small fraction of the degrees of freedom of the entire system, the coupling to the
  rest of the system mimics a coupling to a reservoir.  If the dynamics satisfy this property, the microcanonical, canonical, and grand canonical
  ensembles for the full system all give the correct long-time equilibrium properties of subsystems, the conventional theory of quantum statistical
  mechanics thus applies to the long-time steady states of subsystems, and we say the system thermalizes.
  However, not all closed quantum systems do act as reservoirs that thermalize their subsystems.

  Localized systems, first identified by Anderson \cite{Anderson}, do not act as reservoirs for themselves, and thus do not thermalize.  Instead, the long-time states of subsystems are determined by (and thus can `remember') some local details of the system's initial state.  This is why localized systems are of interest as possible quantum memories.
  The distinction between localization and thermalization is only dynamical. It is invisible if one examines only thermodynamic quantities, which are determined by averaging over an equilibrium ensemble of states and essentially assume thermalization.
    However, the distinction is quite apparent if one looks at the properties of individual exact many-body eigenstates of the system's Hamiltonian.
  Thus we examine the single-eigenstate limits of the microcanonical ensemble, which are able to detect the quantum phase transition (or transitions) between the localized and thermalizing phases.  This suggests a new `eigenstate statistical mechanics' that is very useful in investigating localization, and also reveals a whole new world of localized phases, and quantum phase transitions between them, within the localized regime \cite{LPQO}.

  The bulk of the existing literature about localization focuses on noninteracting systems or on the low temperature limit.
  However, the perturbative arguments of \cite{Fleishman, agkl, Mirlin} and particularly \cite{BAA}, numerical exact diagonalization studies, e.g. \cite{Oganesyan, pal}, and even a recent mathematical proof \cite{IS} have provided strong evidence that localization can occur in highly-excited states of strongly-interacting many-body quantum systems - a phenomenon that has now come to be known as `many-body localization', and has been the subject of considerable recent work \cite{Bauer, Pekker, Vosk, Vosk', Bahri, Chandran, Abanin1, Lbits, Abanin2, Lbits2, MBLMore1, MBLMore2, MBLMore3, MBLMore4, MBLMore5, MBLMore6, MBLMore7, MBLMore8, MBLMore9, MBLMore10, MBLMore11, MBLMore12, MBLMore13, Floquet, Floquet2}.

 Many-body localization (MBL) represents a new frontier of quantum statistical mechanics. Many-body localized systems fail to thermally equilibrate,
 so their long-time states are not captured by conventional equilibrium statistical mechanics.
 At the same time, the existence of interactions allows for a highly non-trivial statistical mechanics of these localized systems \cite{LPQO}. Indeed, many-body localized systems can exhibit a phenomenology that runs counter to theorems of equilibrium statistical mechanics, such as, for example, ordered phases and phase transitions in one-dimensional systems at energies that correspond to high temperatures \cite{LPQO, Bauer, Pekker, Vosk, MBLMore12}.  In this review, we provide a pedagogical introduction to key features of, implications of, and some open questions in many-body localization.  This review is almost entirely about theoretical questions.  Many-body localization as a subject of experimental research is still in its infancy, so a review of that topic would be premature, and we have not attempted it.

  \section{Closed-system many-body quantum mechanics}
  \label{review}

 We are interested in the properties of closed
 many-body quantum systems with a short-range Hamiltonian $H$.  Unless otherwise specified, we consider a time-independent Hamiltonian,
 but below we do occasionally consider the Floquet case of a
 Hamiltonian that is periodic in time.
 By closed we mean that the system is not connected to any `external' environment or to any measuring apparatus.
 Any environment or measuring apparatus that is coupled to the system should instead be treated quantum mechanically, and included as part of the system.
 We assume, unless otherwise specified, that the Hamiltonian $H$ is local in real space, meaning all interactions are short-range.

 Unlike many treatments of quantum condensed matter, we do not focus on the ground state and low-lying excited states, but instead on highly-excited states, which have a non-zero {\it energy density} relative to the ground state even in the thermodynamic limit.  We also do not restrict ourselves to pure states since the closed system may in the past have been coupled to some other degrees of freedom with which it is still entangled. Arbitrary states of a closed quantum system can be treated using the formalism of probability operators (a.k.a. density matrices) \cite{sakurai}.  We work in the Schr\"odinger representation, where the probability operator $\rho(t)$ evolves in time according to
 \begin{equation}
 \rho(t) = e^{- \frac{i H t}{\hbar}} \rho(0) e^{\frac{i H t}{\hbar}}; \qquad i \hbar \frac{d\rho}{dt} = [H, \rho]; \qquad \Tr\{\rho\}=1~.\label{eq: time evolution}
 \end{equation}
 Meanwhile, all other operators $\hat O$ are time-independent, and the expectation value at time $t$ of an observable $O$ with corresponding operator $\hat O$ is
 $\langle\hat O\rangle_t=\Tr\{\hat O \rho(t)\}$.

\begin{figure}
\includegraphics{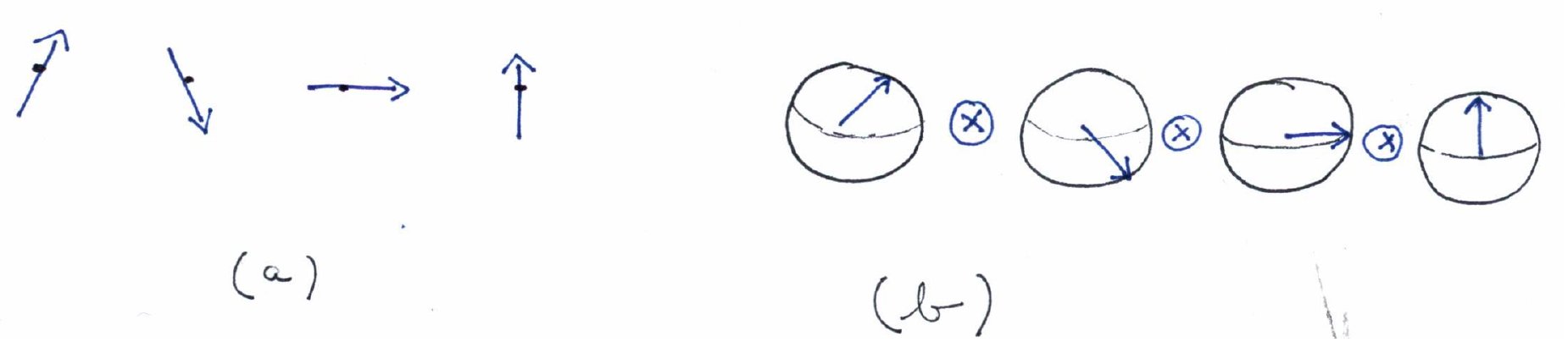}
\caption{The system of interest can be represented as a set of spins on a lattice (a). Each spin has a two dimensional state space, which can be represented on a Bloch sphere (b). The many body pure state space for the full system consists of the outer product of the pure state spaces of each spin, as illustrated in (b).\label{fig1}}
\end{figure}

 For specificity, we now specialize to a quantum system of $N$ two-state systems.  These two-state systems need not be spin-1/2's, but it is convenient to refer to them as spins.  Each spin is located at a point in real space; we consider one, two or three spatial dimensions.  The spins may be randomly located as in \cite{Anderson}, or arrayed on a regular lattice.  The locality that `localization' refers to is in real space.  Each spin has a local space of pure states consisting of two states and all of their complex linear combinations.  And each spin has four linearly-independent operators that can act on it.  These can be represented as $2\times 2$ matrices: the identity matrix $\mathbb{I}_i$ and the three Pauli matrices $\sigma^x_i$, $\sigma^y_i$, $\sigma^z_i$ for spin $i$.  A general mixed state $\rho_i$ of this spin is a linear combination of these operators.  More generally, one could consider a local `spin' with $q$ orthonormal pure states and $q^2$ different operators (including its identity operator).

 The many-body pure-state space of the full system of $N$ spins is the outer product of the pure-state spaces of each spin, Fig. \ref{fig1}.  A convenient (and conventional) basis set for this space is the $2^N$ simultaneous eigenstates of all of the $\{\sigma^z_i\}$.  This basis set has a geometry: it can be represented as the $2^N$ corners of an $N$-dimensional hypercube.  Nearest-neighbor points on this hypercube are states that differ by flipping only one spin.  Each axis of the hypercube represents one spin, so in general this hypercube does not have any rotational symmetry, since each of its axes represents a different point in real space.  The Hamiltonian of the many-body system can be represented as a single quantum particle hopping on this hypercube, with only short-range hopping.  But a many-body state that is localized need not appear localized on this hypercube: For example, a many-spin product pure state that is localized so each spin points in its own particular direction on its own Bloch sphere, with the directions selected randomly and uniformly on the Bloch spheres, appears quite extended when represented on this hypercube.  Such a state will only look localized on the hypercube if the basis choice for each spin corresponds to the state in which that spin is localized.  Note that the geometry of this hypercube is {\it not} the geometry of the space (the `Hilbert space') of the full system's pure states.  The latter space is a $2^N$-dimensional complex vector space, and each of its $2^N$ `axes' is one of the corners of this hypercube.  Thus our system has at least three `spaces' that can be considered: real space, the hypercube of many-body basis states, and the Hilbert space.

 A convenient complete basis for the full system's operators is given by outer-product spin operators of the form, e.g.
 \begin{equation}
 \sigma^x_1 \otimes \sigma^z_2 \otimes \mathbb{I}_3 \otimes...\otimes \sigma^y_N~,
 \end{equation}
 where every spin contributes to the product either its identity operator or one of its Pauli operators.
 Thus the system has $4^N$ linearly-independent operators that can operate on its states and from which one may make its mixed states.
 Again, if we instead have a local `spin' $i$ with $q_i>2$ pure states, it contributes a factor of $q_i^2$ to the number of operators.

Introducing some terminology:  A `$k$-local' operator is an operator of the above form where $k$ of the entries are not identity operators
(i.e., a product operator that acts non-trivially on only $k$ spins).  Such an operator can, for example, be a `hopping' of range $k$ on the hypercube of many-body basis states. Any linear combination of $k$-local operators is also termed a $k$-local operator.
A `global' operator is a $k$-local operator where $k$ is of order $N$.  In contrast, an operator that is local in real space
(henceforth referred to simply as a local operator) is an operator where $k$ is of order one {\it and} the non-identity Pauli operators act only
on spins that are all within distance of order one of each other.

The Hamiltonian $H$ of our system is a sum of local operators.  The system may have a few other extensive conserved quantities that are also sums of local operators.  Examples include spin and particle number.  Like the energy, these quantities can be transported by the system's dynamics.  These extensive conserved operators commute with $H$ in a way that is not `fine-tuned' to the details of $H$.  We do not consider traditional translationally-invariant integrable systems that have an infinite sequence of extensive conserved quantities that are sums of local operators.  Such integrable systems are special cases and are presumably not robust to arbitrary small local changes in $H$.  We note that the `generalized thermalization' of such integrable systems to the `generalized Gibbs ensemble' is an interesting and well-studied subject \cite{GGE0, GGE1, GGE2} that we do not review here.

Any closed many-body quantum system has many conserved quantities that are given by global operators.  A complete linearly-independent set of operators that commute with $H$ consists of each projection operator on to an exact many-body eigenstate of $H$.  These projection operators are all global operators.  But the Hamiltonian itself is a linear combination of these projection operators and is, in contrast, a sum of {\it local} operators (as is also, trivially, the identity operator).  There is a certain sense in which local operators are physical and observable, while global operators are not \cite{lych}:  Another degree of freedom, such as a measuring apparatus, can realistically couple to a quantity that is represented by a local (or $k$-local) operator, while in the thermodynamic limit this is not feasible for a quantity that is represented by a global operator.  Note also that in the thermodynamic limit almost all of this system's full set of $4^N$ operators are global and thus are `unphysical' in this sense.

\section{Quantum thermalization}\label{QT}

\subsection{What is thermalization?}\label{what}

We are now ready to discuss the {\it apparent} paradox of quantum thermalization.  A quantum system in thermal equilibrium is fully characterized by a small number of parameters (temperature, chemical potential, etc.: one parameter for each extensive conserved quantity), suggesting that the process of going to thermal equilibrium is associated with the `erasure' of the system's `memory' of all other details about its initial state.  However, unitary time evolution cannot erase information, and thus all quantum information about the initial state must be preserved within the (closed) system for all times.

\begin{figure}
\includegraphics{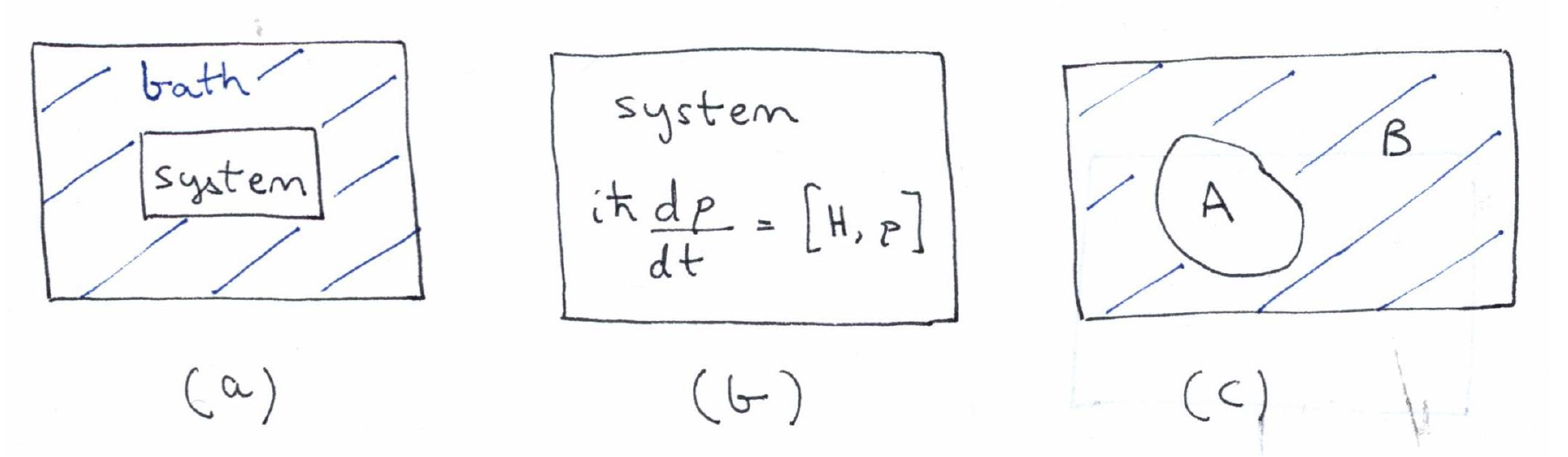}
\caption{\label{fig2} (a) Conventional quantum statistical mechanics assumes that the system of interest is coupled to a reservoir (or bath), with which it can exchange energy and particles. (b) Here we are interested in the statistical mechanics of a closed quantum system undergoing unitary time evolution. There is no external reservoir. (c) It can be useful to partition the closed quantum system into a subsystem (A) and `everything else' (B). If the system quantum thermalizes, then the region (B) is able to act as a bath for the subsystem (A).}
\end{figure}

The resolution to this apparent paradox is that the memory of the local properties of the system's initial state is not erased by unitary time evolution,
but is instead `hidden' if the system thermalizes.  Spreading of quantum entanglement moves the information about the initial state so that at long time
it is inaccessible, since recovering that information would require measuring global operators.  This is the process of `decoherence'.
In particular, if we restrict to a subsystem which is a small fraction of the full system, then thermalization means that at long times the state of this subsystem
is as if it were in thermal equilibrium in contact with a reservoir characterized by a temperature $T$, a chemical potential $\mu$, etc., since in fact it is,
with the reservoir being the remainder of our closed system (Fig. \ref{fig2}).  It is this ability of quantum systems to act as reservoirs for their subsystems
that underpins equilibrium quantum statistical mechanics.

We now provide a somewhat more precise description of quantum thermalization.  To keep the discussion as simple as possible, we consider a closed system that does not have any extensive conserved quantities other than energy, so that if it thermalizes, the thermal state is described by one parameter, the temperature.  A generalization to systems with a few more conserved quantities does not need substantial additional concepts that are not present in this simpler case.  The interactions in the system's Hamiltonian must `connect' all of its degrees of freedom, so the system does not contain any subsystems that are themselves isolated closed subsystems not in contact with the remainder of the system.
We partition the full quantum system into a subsystem A and its `environment' B,
which contains all the degrees of freedom not in A.  We will need to take the thermodynamic limit on the number of degrees of freedom in B,
such that in this limit the fraction of the full system's degrees of freedom that are in A goes to zero.
Any choice of subsystem A is acceptable, as long as the degrees of freedom within A are defined by $k$-local operators with finite $k$.
A concrete case can be that A is a fixed compact subregion (in real space).  However, different choices, such as a subsystem in momentum space, or a set of degrees of freedom that are well separated in real space, or even a single degree of freedom also constitute acceptable subsystems. For clarity, we consider the case where $A$ is a fixed compact subregion of the full system, and B is taken to infinite volume by adding more degrees of
freedom that are in the limit arbitrarily far from A. Thus as we take the thermodynamic limit on B, we have a sequence of systems and their Hamiltonians, with the number of degrees of freedom increasing without limit.  In the concrete case mentioned, let the changes to $H$ as the system's size is increased be only at the locations far from A where the new degrees of freedom are being added to B.

For each system in this sequence as we take the thermodynamic limit we need to consider a set of initial states $\rho(t=0)$.  Let's consider initial states that
if they do thermalize will thermalize to a given temperature $T$.  Each system in our sequence has an equilibrium expectation value of the total energy
$\langle H\rangle_T$ at temperature $T$.  We consider a sequence of initial states such that the mean-square deviation within each initial state of the total energy of the system from
$\langle H\rangle_T$ grows no faster than the volume of the full system.  Although we thus fix the average energy density to be at its equilibrium value,
we do not otherwise constrain how this energy is initially distributed, and we are particularly interested in initial states where that initial distribution of
energy is far from equilibrium, since these are a type of initial state that may fail to thermalize.
Each system and each initial state is then time-evolved according to (\ref{eq: time evolution}).  The probability operator $\rho_A(t)$ (a.k.a. reduced density matrix) of subsystem A at time $t$ is obtained from $\rho(t)$ of the full system by taking a partial trace over all of the degrees of freedom in B: $\rho_A(t) = \Tr_B\{\rho(t)\}$.  The same system at equilibrium at temperature $T$ has Boltzmann probability operator $\rho^{(eq)}(T)=Z^{-1}(T)\exp{(-H/k_BT)}$ for the full system and thus $\rho_A^{(eq)}(T) = \Tr_B\{\rho^{(eq)}(T)\}$ for the subsystem.  The system thermalizes for this temperature if in the long-time and large-system limit $\rho_A(t)=\rho_A^{(eq)}(T)$ for all subsystems A \cite{Deutsch, Srednicki,tasaki, Rigol}.  These two limits must be taken together:  For a finite system the dynamics is quasiperiodic so $\rho(t)$ does not have a long-time limit, while for finite time the diffusive transport in a thermalizing system only reaches a finite distance.

Thermalization is of particular interest for initial states that are well out of equilibrium.  These are atypical initial states, since at equilibrium the system is in typical states.  Thus when we say a system thermalizes, for this to be an interesting statement it must certainly apply to {\it some} atypical initial states.  The conventional story in equilibrium statistical mechanics is that {\it all} out-of-equilibrium initial states at this energy will thermalize in the limit of infinite time to equilibrium at the corresponding temperature.  Thus it seems reasonable to assume that if a system does thermalize for a given temperature, this means {\it all} initial states at that energy thermalize.  This is a strong statement with some strong consequences.  It seems to be well beyond what can be proved for any generic system, but it seems to be consistent with what we know so far about systems that do thermalize.

It is interesting to consider a many-body Floquet system, a system with a time-dependent Hamiltonian that is periodic in time with period $\tau$, so $H(t)=H(t+\tau)$.
In such systems energy is only conserved modulo $2\pi\hbar/\tau$, so there is no conserved energy {\it density}.  Let us consider a Floquet system that has no
conserved densities at all, and thus nothing to transport.  A periodically driven system with an unbounded Hilbert space cannot really thermalize, instead it can absorb energy from the periodic drive without limit.  However, when a periodically driven system with a bounded Hilbert space thermalizes, it thermalizes to infinite temperature \cite{Floquet2}, at which point all pure states of any small subsystem are equiprobable.
When such a system thermalizes, it still serves as a reservoir for its subsystems.  But a reservoir of what,
since there is no conserved quantity that is being exchanged between the subsystem and the reservoir?  This suggests that the essential function of a reservoir is
not as a source and sink of energy, particles, or other such conserved quantities.  Instead the most basic function of the reservoir may be to provide other quantum
degrees of freedom that the subsystem gets so entangled with that no information about the initial state of the subsystem remains locally observable.

\subsection{The Eigenstate Thermalization Hypothesis}\label{ETH}

If a system at a given temperature does indeed thermalize for every such initial state $\rho(t=0)$, then it is quite instructive to consider initializing
the system in a pure state that is one of the many-body eigenstates of $H$.  The time evolution of the system then becomes trivial: $\rho(t) = \rho(0)$,
so thermalization of all initial states implies that {\it all many-body eigenstates of $H$ are thermal.}
This statement is known as the {\it Eigenstate Thermalization Hypothesis} (ETH) \cite{Deutsch, Srednicki,tasaki, Rigol}.
Before we discuss the ETH in more detail, we must emphasize that we are not considering the exact eigenstates of $H$ because these are realistic states of a many-body system.  On the contrary, they are impossible to prepare in the laboratory as initial states.  The initial states of a many-body system that can actually be prepared are not even pure states, much less eigenstates.  As always, the focus on eigenstates is instead because they are an essential {\it tool} in understanding the dynamics.  When written as a `density matrix' using the eigenbasis $\{|n\rangle\}$ of $H$, the dynamics of $\rho(t)$ is simple: its diagonal terms $\rho_{nn}=\langle n|\rho|n\rangle$ are constant, while its off-diagonal terms $\rho_{nm}=\langle n|\rho|m\rangle$ each simply `precess' in the complex plane at a constant rate given by the difference in energy between the two eigenstates involved: $\rho_{nm}(t)=\rho_{nm}(0)\exp{(i(E_m-E_n)t/\hbar)}$.

To define the ETH a little more precisely, consider an eigenstate $H|n\rangle=E_n|n\rangle$.  Its energy $E_n$ is the thermal equilibrium energy at temperature $T_n$, so $E_n=\langle H\rangle_{T_n}$.  If the full system is in this eigenstate, then $\rho=\rho^{(n)}=|n\rangle\langle n|$, and thus
$\rho_A^{(n)}=\Tr_B\{|n\rangle\langle n|\}$ is the state of subsystem A.  The ETH asserts that in the thermodynamic limit the subsystem is at thermal equilibrium: $\rho_A^{(n)}=\rho_A^{(eq)}(T_n)$.  One noteworthy consequence of this is that the entropy of entanglement,  $S_{AB}=-k_B\Tr_A\{\rho_A^{(n)}\log\rho_A^{(n)}\}$, between A and B in this eigenstate of the full system is equal to the equilibrium thermal entropy of the (smaller) subsystem A.  For eigenstates with temperature $T_n \neq 0$, this entropy is proportional to the volume of A.  Thus the entanglement entropy in thermal eigenstates obeys {\it `volume-law'} scaling, as it does also in any thermal pure state of the full system.

Another requirement of the ETH is that the matrix elements of
 subsystem operators between distinct eigenstates vanish strongly enough in the thermodynamic limit
\cite{Rigol, ETH4}.  This is needed to ensure that temporal fluctuations of $\rho_A(t)$ vanish
(as opposed to the weaker scenario in which only the time-average of $\rho_A(t)$ is thermal).

For many-body Floquet systems
the dynamically stable many-body states of the full system that play the role of the Hamiltonian's eigenstates are the eigenstates of $U(\tau)$, the unitary operator
that takes the full system forward in time by one period $\tau$.  Of course these eigenstates of $U(\tau)$ are not completely stationary, they only return to
the same state after a full period of the drive.  For a system with no conserved densities, the equilibrium state of a subsystem has $\rho_A^{(eq)}$ simply equal
to the identity operator on that subsystem (times a normalizing factor), so all %
pure states of the subsystem are equally probable.   Thus when the ETH applies to such Floquet systems, it means that in the thermodynamic limit all eigenstates of the full system give equal probabilities for all possible pure states of any finite subsystem.

The ETH is an {\it hypothesis}.  It is not true for one broad class of systems, namely those that are many-body Anderson localized, as we discuss below.
For systems where the ETH appears to be true, this is difficult to thoroughly test numerically, since it requires obtaining the many-body
eigenstates of the system's Hamiltonian from exact diagonalization, and extrapolating to the thermodynamic limit.
In order to have thermalization of all initial states that can realistically be prepared, it seems that one does not need the ETH to be true for absolutely {\it all} eigenstates; a weaker {\it almost all} should suffice.  And if we look at the numerical results, they do strongly support the proposal that there are systems where at least {\it almost all} eigenstates obey the ETH \cite{Rigol, pal, ETH1, ETH2, ETH3, ETH4, ETH5, ETH6, ETH7, ETH7.5, ETH8, ETH9, ETH10}.  But the simpler and it seems more plausible scenario is that if the ETH is true for a given system at a given temperature, then it is true there for {\it all} eigenstates.  If there were certain rare eigenstates that violate the ETH even though they are essentially degenerate with the typical eigenstates that obey the ETH, we would need to understand what is so special about these rare eigenstates.

We note that if the ETH is true and the state $\rho$ of the system is
diagonal in the energy eigenbasis, then all subsystems are at thermal equilibrium. This prompts the question: when the ETH is true, how does one construct out-of-equilibrium states?
The answer is that out-of-equilibrium states have special structure off the diagonal when $\rho$ is written in the energy eigenbasis.  Thus they have special coherence patterns between eigenstates of different energies.
Quantum thermalization of an out-of-equilibrium initial state requires that the contributions of this off-diagonal coherence to local observables must vanish
at long times. This happens due to {\it dephasing}: while the diagonal terms in $\rho$ (in the energy eigenbasis) are time-independent, the off-diagonal terms
have phase oscillations at frequencies set by the energy differences between the corresponding eigenstates.  Thus unitary time evolution `scrambles' the phases
of the off-diagonal terms in $\rho$, such that at long times and in the thermodynamic limit their contributions to any local observables come with
effectively random phases, and thus cancel.  In this sense, we see that in the energy eigenbasis equilibration is `simply' dephasing.

The ETH motivates introducing a new set of ensembles to use in quantum statistical mechanics, namely the {\it single-eigenstate ensembles} that each consist of a single eigenstate of the full system's Hamiltonian.  When the ETH is true, these ensembles all give the correct thermal equilibrium properties of subsystems, just like the traditional statistical mechanical ensembles.  The single-eigenstate ensembles may be viewed as the limiting case of the microcanonical ensemble where the energy window has been reduced to the limit where it contains only one eigenstate.  The full payoff of introducing these new ensembles and the resulting `single-eigenstate statistical mechanics' becomes clear when one considers systems that do not obey the ETH, as we will discuss below.

While the ETH and quantum thermalization appear to apply to a large class of closed quantum systems, not all systems quantum thermalize.  One well-known exception is traditional integrable systems, which possess an infinite set of extensive conserved quantities.  It has, however, been argued that such integrable systems exhibit their own version of quantum thermalization, to a `generalized Gibbs ensemble' (GGE), and they have their own version of the ETH \cite{GGE0, GGE1, GGE2}. The focus of the remainder of this review, however, is on a class of systems which fail to quantum thermalize in any sense, and where the many-body eigenstates violate the ETH.  These are Anderson-localized systems.

\section{Localized systems}
The concept of localization was first introduced by Anderson \cite{Anderson}, and applies to systems with quenched disorder (see, however, Sec.\ref{annealed} for a discussion of the possibility of many-body localization in lattice systems with translationally-invariant Hamiltonians).  The label `localization' is used for at least three different situations:  Most experimental work on localization to date is about regimes near ground-state quantum phase transitions (e.g., metal-insulator transitions).  This is not a topic of this review.  Most theory work on localization to date is concerned with non-interacting particles in a random potential (or waves in random media without nonlinearities).  We will briefly review this work in Sec. \ref{single particle localization}, but it is also not a main topic of this review.  The focus of this review is on {\it interacting} many-body systems, and we are interested not in ground states or the low-energy limit, but instead in highly-excited states of such systems at energies that would correspond to nonzero temperature if the system thermalized.  The discussion of many-body localization begins in Sec. \ref{mbl}, and occupies the remainder of this review.

  \subsection{Single-particle localization}
  \label{single particle localization}
We briefly review single-particle localization (for a more complete discussion, see e.g. \cite{localizationRMP}).  The essential physics of single-particle localization
can be illustrated with a tight-binding model of a single quantum particle hopping on an infinite lattice,
with Hamiltonian
  \begin{equation}
 H =t \sum_{\langle ij \rangle} (c^{\dag}_i c_j + c^{\dag}_j c_i) + \sum_{i} U_i c^{\dag}_i c_i   ~,
 \end{equation}
 where $U_i$ is a static random onsite potential, $t\neq 0$ is a nearest-neighbor hopping, and $c^{\dag}_i$ creates a particle on the site $i$.  Consider the motion of a single particle in this system.  In three or more dimensions, and for weak enough disorder, the eigenstates of this Hamiltonian can be `extended' with weight on all sites, and diffusive dynamics of a particle initialized in a wave packet composed of such extended eigenstates.  However, in one or two dimensions, (and in three or more dimensions with strong enough disorder), the eigenstates are all exponentially localized, with wavefunctions that have the asymptotic long distance form $\psi_{\alpha}(\vec{r}) \sim \exp\left(- \frac{|\vec{r} - \vec{R}_{\alpha}|}{\xi}\right)$, where $\xi$ is the localization length, which depends on the disorder strength and on the energy.  This state $\alpha$ is localized near position $\vec{R}_{\alpha}$, and a particle remains localized near the location where it is initially introduced.  In three or more dimensions, the transition between localized and extended states happens via special `critical' states at the `mobility edge', which display power-law localization.

  A straightforward basis transformation recasts this Hamiltonian into the simple form
  \begin{equation}
  H = \sum_{\alpha} E_{\alpha} c^{\dag}_{\alpha} c_{\alpha} ~,
  \end{equation}
  where $c^{\dag}_{\alpha}$ creates a particle in the single-particle eigenstate $|\alpha\rangle$, and $E_{\alpha}$ is the eigenenergy of a particle occupying this state.
  We have written this Hamiltonian in second-quantized form, so it is also a many-particle Hamiltonian, although still without interactions between the particles.  Its many-particle eigenstates are simple product
  states in terms of the single-particle eigenstates, and can be labeled by the occupation numbers of all the various single-particle eigenstates $|\alpha\rangle$.
  For systems where at least some of the single-particle eigenstates are localized, almost all of these many-particle eigenstates
  violate the ETH.  As an example to show the lack of quantum thermalization, initialize the system with a spatially non-uniform density of particles in the localized states over a large length scale: this non-uniform initial density pattern then survives for all times, since those particles are localized.

While particles hopping in a random potential provide the best known example of single-particle localization, the phenomenon is more general,
and occurs also in systems of spins.  For example, a spin-1/2 version of single-particle localization arises in a system governed by the spin Hamiltonian
  \begin{equation}
  H = \sum_i h_i \sigma^z_i + \sum_{ij} J_{ij} \vec\sigma_i \cdot \vec\sigma_j ~. \label{eq: model2}
  \end{equation}
The onsite magnetic fields $h_i$ are static random variables, e.g. taken from a continuous probability distribution of width $W>0$, and the spin `hoppings' and interactions $J_{ij}$ are
strictly short range in real space (for specificity, take $J_{ij} = J \neq 0$ for nearest neighbors and zero otherwise).  One then considers an initial condition
where there is a single spin `up' (let us say, on site $i$) with all other spins down, and asks about the dynamics of this up spin.

If there are any localized single-particle states with weight at site $i$, there is a non-zero probability that the site $i$ still hosts an up spin even at infinite time - i.e. that memory of the initial conditions is preserved in a local observable for infinite times.  This manifestly constitutes a failure of quantum thermalization, and is quite similar to the context in which single-particle localization was first established in Ref. \cite{Anderson}.

  \subsection{Many-body localization}
  \label{mbl}
 We now turn to many-body localization (MBL): localization with interactions.  This physics can be studied in models with mobile particles.  Since it occurs at high
 energies, it appears that there are no major differences between MBL of bosons and fermions.
 But
  the physics of the MBL phenomenon is most simply exposed in the context of spin models.  The model above (\ref{eq: model2}) is fine to illustrate MBL; we simply
  consider states where the densities of up and down spins are both nonzero, instead of considering only one flipped spin.
We are interested in whether or not a system governed by the above Hamiltonian (\ref{eq: model2}) quantum thermalizes for arbitrary initial conditions.  We therefore ask whether the {\it many-body} eigenstates of the above Hamiltonian obey the ETH, since this is a necessary condition for quantum thermalization.

At $J=0$, the many-body eigenstates of (\ref{eq: model2}) are simply product states of the form $|\sigma^z_1\rangle \otimes |\sigma^z_2\rangle \otimes ...$,
and the system is fully localized.  For nonzero $J$, in the regime $J \ll W$ one can construct the many-body eigenstates perturbatively in small $J$ \cite{Fleishman, agkl, Mirlin, BAA}.
Since in this regime the typical level splittings between nearest-neighbor sites are much larger than the interactions $J$, the states on different sites are
typically only weakly hybridized.  This line of argument (similar to that employed by Anderson in \cite{Anderson}) leads one to conclude that for sufficiently
strong disorder $W \gg J$, DC spin transport and energy transport is absent, and quantum thermalization therefore does not occur at any order in perturbation theory \cite{BAA}.
While this argument is perturbative, and is limited to the weak interaction regime, extensive numerical evidence (mostly for one-dimensional
systems \cite{Oganesyan, pal, Bauer}) suggests that while the high-entropy eigenstates of (\ref{eq: model2}) do obey the ETH for weak disorder (small $W$),
for strong enough disorder all of the eigenstates violate ETH.  Moreover, the violation of ETH apparently occurs even for {\it strong} interactions,
outside the regime where the perturbation theory can be controlled.

There is a quantum phase transition as one varies the disorder strength or the energy density
between the thermal phase in which we expect all the eigenstates obey the ETH and the system quantum thermalizes,
and the `many-body localized' phase wherein all the eigenstates do {\it not} obey
the ETH, and some memory of the local initial conditions can survive in local observables for arbitrarily long times.
Many questions about the nature of this phase transition remain open (see Sec.\ref{sec: phase transition}).  This transition is an {\it eigenstate} phase transition, marked by a sharp change in properties of the many-body eigenstates and thus in the dynamics of the system, so this transition is visible if one studies the system using the single-eigenstate ensembles.  However, this transition is invisible to equilibrium thermodynamics and to the traditional statistical mechanical ensembles, since they average over many eigenstates.  Indeed, the many-body localization transition marks the breakdown of the applicability of equilibrium quantum statistical mechanics to the system's long-time properties.

Although localization is usually discussed for systems with static randomness, it has long been known that nonrandom systems with instead quasiperiodicity can support single-particle localization.  This has recently been demonstrated \cite{Iyer} to also remain true for many-body localization.

  While the original idea of localization came from considering spin systems \cite{Anderson}, and spin models provide simple examples in which to explore MBL,
  important more recent work about MBL considered also systems of fermions:  In \cite{agkl} it was pointed out that a system of interacting fermions in zero dimensions
  (a `quantum dot') can be approximated by a single-particle localization problem on a Cayley tree.  This result was exploited in \cite{BAA, Mirlin} to show that
  single-particle localization in spatial dimensions $d \ge 1$ is robust to weak nonzero interactions, to all orders in perturbation theory.
  A feature that can occur that was emphasized in Ref. \cite{BAA} is
  a {\it many-body mobility edge} at an extensive energy in systems where some but not all many-body eigenstates are localized.  The usual behavior in the energy regime that corresponds to positive temperature is that all many-body eigenstates with energy density above the mobility edge (thus in the {\it thermal phase}) obey the ETH, whereas all eigenstates with energy density below the mobility edge are in the localized phase.
  Models with an `inverted' mobility edge can also be constructed, where localization occurs only for those eigenstates above a critical energy density, by making use of a model where the single particle level spacing increases with energy \cite{MBLMore11}.  We are not aware of any models exhibiting multiple mobility edges in the positive temperature range, but there does not seem to be any obvious reason why such models cannot be constructed.  The behavior of MBL systems with a mobility edge is more complex than that of systems where all eigenstates are localized.  In particular, one may worry about rare regions within a localized state that have a local energy density close to that of the mobility edge (a new type of quantum Griffiths phenomenon).

We close this Section by emphasizing one key feature of the MBL phenomenon:  Closed MBL systems do not quantum thermalize, so some memory of the local initial conditions is preserved in local observables for arbitrarily long times.
This implies that the DC conductivity of any conserved densities must be strictly zero in the MBL phase.  While the vanishing of DC transport is
a useful diagnostic for MBL in certain systems, it is {\it not} the key distinguishing feature.
Indeed, one can consider a many-body Floquet system with {\it no} conserved
local densities, so it has no meaningful DC transport properties.  But still such Floquet systems may have a quantum phase transition between thermal and MBL phases  (see e.g. \cite{Floquet, Floquet2}).

 \begin{table}
 \begin{tabular}{c|c|c}
	Thermal phase & Single-particle localized & Many-body localized\\
		\hline
	Memory of initial conditions & Some memory of local initial & Some memory of local initial \\ `hidden' in global operators & conditions preserved in local & conditions preserved in local \\ at long times & observables at long times & observables at long times. \\
	\hline
	ETH true & ETH false & ETH false\\
	\hline
	May have non-zero DC conductivity & Zero DC conductivity & Zero DC conductivity\\
	\hline
	Continuous local spectrum  & Discrete local spectrum & Discrete local spectrum  \\
	\hline
	Eigenstates with & Eigenstates with & Eigenstates with \\
	volume-law entanglement & area-law entanglement & area-law entanglement \\
	\hline
	Power-law spreading of entanglement & No spreading of entanglement & Logarithmic spreading of entanglement \\
	from non-entangled initial condition &   & from non-entangled initial condition \\
	\hline
	Dephasing and dissipation & No dephasing, no dissipation & Dephasing but no dissipation\\

  \end{tabular}
   \caption{\label{table of contrasts} A list of some properties of the many-body-localized phase, contrasted with properties of the thermal and the single-particle-localized phases. The spreading of entanglement is discussed further in Sec.IV-C. Local spectra are discussed further in Sec.IV-D.}
  \end{table}

  \subsection{A phenomenology of many-body localized systems}
  \label{phenomenology}

  We begin this Section by tabulating some properties of the thermal (non-localized), single-particle localized, and many-body localized phases (Table \ref{table of contrasts}).  The first three lines of this table follow from the discussion in Sec. \ref{mbl}.  However, explaining the rest of this table requires a little more discussion, which we now provide.

Let us assume for specificity that we have a system of $N$ local two-state degrees of freedom $\{\vec{\sigma}_i \}$, which we refer to as the `p-bits' (p=physical).
These could be the spins from (\ref{eq: model2}), or could be e.g. the occupation numbers of localized single-particle orbitals in a system of fermions in a
random potential.  An analogous argument can be constructed for objects with more than two states, but we stick to this two-state example for specificity.
Let us further assume that the p-bits are governed by a Hamiltonian with quenched disorder and strictly short-range interactions.
For strong enough disorder, such a Hamiltonian can be in the fully many-body localized (FMBL) regime,
wherein all the many-body eigenstates of the Hamiltonian are localized.
It was argued in \cite{Lbits, Abanin2, Abanin1, Lbits2} that in this FMBL regime, one can define a set of localized two-state degrees of freedom, with Pauli operators
$\{\vec{\tau}_i\}$, henceforth called `l-bits' (l=localized) such that the Hamiltonian when written in terms of these new variables takes the form
\begin{equation}
H = E_0 + \sum_{i} \tau^z_i + \sum_{ij} J_{ij} \tau^z_i \tau^z_j + \sum_{n=1}^{\infty} \sum_{i,j,\{k\}} K^{(n)}_{i \{k\} j } \tau^z_i \tau^z_{k_1}...\tau^z_{k_n} \tau^z_j ~, \label{eq: lbits}
\end{equation}
where the sums are restricted so that each term appears only once, and $E_0$ is some constant energy offset which may be zero and which has no relevance for the
closed system's dynamics.  The typical magnitudes of the interactions $J_{ij}$ and $K^{(n)}_{i \{k\} j }$ fall off exponentially with distance,
as do their probabilities of being large.

The intuition underlying this `l-bit Hamiltonian' (\ref{eq: lbits}) is that in the localized phase, since there is no transport, there should be a set of localized conserved charges which are constants of motion of the system. For example, for noninteracting particles moving in a disordered potential, these constants of motion would be the occupation numbers of the localized single-particle orbitals.  For weakly interacting systems, the l-bits $\vec{\tau}_i$ should have substantial overlap with the `bare' p-bits $\vec{\sigma}_i$, and indeed may be viewed as `dressed' p-bits, with a `dressing' that falls off exponentially in real space. The existence of long-range interactions between l-bits follows from the fact that although the Hamiltonian only couples p-bits that are nearby in real space, each of those p-bits has non-zero (but typically exponentially small in the distance) weight on distant l-bits.

 One appealing approach to constructing l-bits is to start with p-bits, and then add the appropriate dressing to make operators that commute with each other and
 with the Hamiltonian order by order in perturbation theory in the p-bit interactions.  This approach will fail to give a unique result away from the limit of weak interactions,
 and also for some l-bits even in the weakly interacting regime due to the appearance of `resonances' (degeneracies in the perturbation theory)
 which make the definition of the l-bits ambiguous. Nevertheless, it was argued in \cite{Lbits, Abanin1, Abanin2, Lbits2} that localized l-bits do exist for FMBL systems, and each such l-bit has overlap with distant p-bits that is typically exponentially
 small in the distance.  Something essentially equivalent to this is proven in \cite{IS}.
 No such construction is possible in the thermal phase.  Whether any such construction is possible in the localized phase for MBL systems with a many-body mobility edge
 remains an open question.

    The l-bit Hamiltonian (\ref{eq: lbits}) provides a useful tool for describing various properties of FMBL systems.  The eigenstates of the Hamiltonian (\ref{eq: lbits}) are simply the simultaneous eigenstates of all the $\{\tau^z_i\}$.  In a generic state the dynamics of a single l-bit are in a certain sense trivial - each l-bit precesses about its $z$ axis at a rate set by its interactions with all other $\{\tau^z_i\}$.  In a generic state where all the $\{\tau^z_i\}$ are uncertain, this precession of l-bits produces entanglement and dephasing.
    Nevertheless, dynamically there are no `flips' of the $\{\tau^z_i\}$, and thus no `dissipation'.  As a result, the dephasing can in principle be reversed by spin echo procedures, so MBL systems can in principle be used to store and retrieve quantum information.  However, to fully reverse the dephasing of a particular l-bit, we need to be able to manipulate only that l-bit, whereas in a general experiment all that one has access to are the p-bits.  Recent work \cite{Serbynetal} suggests that high fidelity spin echo measurements are indeed possible even when one has access to only p-bits, at least in the `perturbative' regime where the typical l-bits are weakly dressed p-bits.  How strong a spin echo signal can be obtained by doing the echo procedure on a bare p-bit away from the perturbative regime remains an open question.

    The structure of the Hamiltonian (\ref{eq: lbits}) can also be used to understand \cite{Abanin1,Lbits} the logarithmic spreading of entanglement in the FMBL phase, observed numerically in \cite{MBLMore1,Bardarson}. It is useful to first consider how entanglement spreading works in thermalizing and in single-particle localized systems, and to contrast these with many-body localized systems:
    In thermalizing systems, the interaction of two p-bits $A$ and $B$ generically causes them to become entangled with each other. The subsequent interaction of bits $B$ and $C$ generically produces entanglement not just between $B$ and $C$, but also between $A$ and $C$.  As a result, entanglement spreads ballistically in some systems \cite{kim,vah}, with a speed akin to a Lieb-Robinson velocity \cite{Lieb}.  In contrast, in single-particle localized systems, there are no interactions between l-bits, so the dynamics does not generate any entanglement between l-bits.

\begin{figure}
\includegraphics{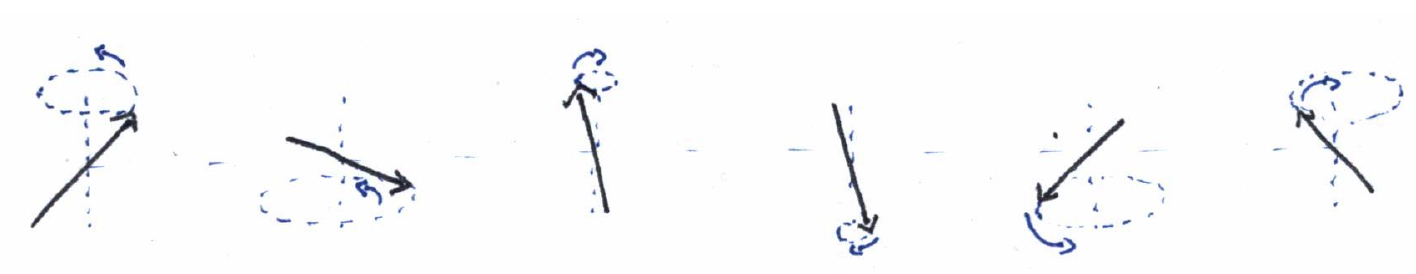}
\caption{\label{precession} An illustration of the dynamical behavior of the l-bits. The l-bits are a set of spins, whose $z$ component does not change, but which precess about the $z$ axis at a rate determined by the effective interactions with all other l-bits. This picture can be used to understand e.g. the logarithmic spreading of entanglement in the fully many body localized phase. }
\end{figure}

    The entanglement spreading in FMBL systems is intermediate between thermalizing and non-interacting localized systems.  Entanglement spreading does occur, because the system is interacting.  However, two l-bits can get entangled only by their direct interaction, and not through a mutual interaction with a third l-bit.  This is because the interaction of $A$ with $B$ depends only on the $\tau^z$ value for l-bit $B$, and the $\tau^z$ value for l-bit $B$ is unaffected by its interaction with a third l-bit $C$, being a constant of motion.  In other words, l-bits become entangled only through their direct interaction, because the l-bit Hamiltonian (\ref{eq: lbits}) has no dissipation (no spin flips).

 To quantify the interaction between l-bits, we define the effective interaction between l-bits $i$ and $j$ as
    \begin{equation}
  J^{eff}_{ij} =  J_{ij} + \sum_{n=1}^{\infty} \sum_{\{k\}} K^{(n)}_{i \{k\} j }  \tau^z_{k_1}...\tau^z_{k_n} ~. \label{eq: Jeff}
    \end{equation}
    Note that this interaction depends on all the other $\{\tau^z_k\}$.  In the localized phase we expect this interaction to typically fall off with the distance $L$ between the two l-bits as $J^{eff}\sim J_0\exp{(-L/\xi)}$.  If these two l-bits are initially not entangled, this interaction will entangle them after a time $t$ such that $J^{eff}t\ge 1$.  As a result, after a time $t$, the system's dynamics produces entanglement between all l-bits within a distance $L \sim \xi \ln (J_0t)$ of each other.  This is the origin of the logarithmic growth of entanglement with time within the MBL phase.

    Note that there are exponentially many multispin terms contributing to the effective interaction (\ref{eq: Jeff}), and only a single two-spin term.
    Thus, for long distances the effective interaction is dominated by the multispin terms, and the individual multispin terms in the sum
    (\ref{eq: Jeff}) typically fall off exponentially with a decay length much less than $\xi$.  The dominant multispin terms have all the $k_i$ near
    the straight line between sites $i$ and $j$.  Flipping a single spin between $i$ and $j$ changes the sign of an appreciable fraction of the multispin terms,
    such that the effective interactions change dramatically from one many-body eigenstate to the next - a form of `temperature chaos' analogous to spin glasses \cite{bm}.
    The effective interaction relevant for a particular experiment is obtained by taking the appropriate average of (\ref{eq: Jeff}) over the particular
    state in which the system is prepared: this is discussed more fully in \cite{Lbits2}.
    The effective interaction can in principle be measured experimentally through a double electron-electron resonance type protocol, as discussed in \cite{Serbynetal}.

    The l-bit phenomenology introduced above works for systems that are {\it fully} many-body localized, i.e. all the many-body eigenstates of the Hamiltonian are localized.  Whether an analogous construction exists for Hamiltonians that have both extended and localized eigenstates, separated in energy by a many-body mobility edge, remains an open question.  Naively, one might think that a similar construction, restricted to localized states with energies below the mobility edge, might have a chance of success.  However, the proper treatment of rare regions where the local energy density approaches the many-body mobility edge (a new type of Griffiths phenomenon) complicates this line of reasoning, such that the development of an analogous phenomenology for systems with a many-body mobility edge remains an open problem.

\subsection{Open systems, local spectra}
\label{open systems}
\begin{figure}
\includegraphics{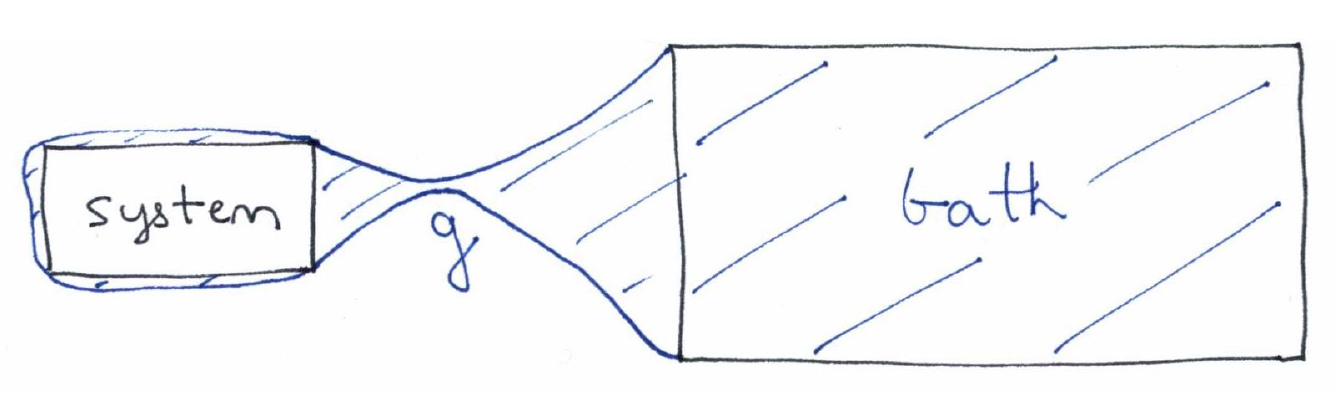}
\caption{\label{fig4} One can consider an almost perfectly isolated quantum system, i.e. a quantum system that is coupled to an external bath with a
weak but non-zero coupling $g$ that couples (weakly) to all of the degrees of freedom in the system.}
\end{figure}

While perfectly isolated MBL systems are a useful theoretical idealization,
any actual physical system always has some small coupling to its external environment (a `bath').
 We therefore are interested in not just the limit of perfect isolation, but also how this limit is approached, and what aspects of
 MBL phenomena survive in the presence of imperfect isolation from the bath (Fig. \ref{fig4}).  Provided that the bath is large enough and
 itself thermalizes, the exact eigenstates of the coupled system and bath for nonzero coupling will generically obey the ETH.
 However, localization of the system is recovered in the limit when the coupling to the bath is taken to zero.  This situation was explored
 in \cite{spectral}, which considered a MBL system coupled to a bath.

Some quite instructive `probes' of localization behavior are the frequency-dependent spectral functions $A(\omega)$ of local operators of the system \cite{spectral}.
If the system is localized and isolated, and (unrealistically) is in an exact many-body eigenstate of the system's Hamiltonian, then the local spectra consist of
discrete delta-functions, each of which arises from flipping some set of the local l-bits near the location being probed.  But the actual state of a real physical
system will not be pure, instead it is a mixed state that populates a set of eigenstates with extensive entropy.  If the system is at equilibrium with a
thermal bath then this mixed state is the Boltzmann-Gibbs distribution.  In these latter cases, for an interacting MBL system the number of delta functions in a
local spectral function of the isolated system is exponentially large in the system's volume, because the spectral lines change in frequency between eigenstates due
to the interactions between l-bits.
The most robust spectral signature of MBL is a `soft gap' at zero frequency in the spectra of local operators, due to level repulsion between the local excitations \cite{spectral, spectralnumerics}.
The local spectral function of a specific isolated system at a specific location may also show gaps that are peculiar to that location; gaps that are thus not present in spatially-averaged spectral functions.
In isolated one-dimensional systems with short enough localization length (referred to in \cite{spectral} as systems exhibiting `strong MBL'), there are spectral gaps
at a infinite hierarchy of scales, with the local spectrum being discrete and having a structure analogous to a Cantor set.  In more than one dimension,
or in one-dimensional systems with longer localization length (referred to in \cite{spectral} as systems exhibiting `weak MBL'), the discrete lines in the local
spectrum of a mixed state fill in to make a piecewise continuous spectrum, which has only a finite number of gaps
(including the `soft gap' at zero frequency which is always present in isolated MBL systems).

Introducing a non-zero coupling to a thermal bath causes the spectral lines to broaden into Lorentzians, erasing all structure in the local spectral functions on
energy scales smaller than the Lorentzian width.  The Lorentzian width goes continuously to zero in the limit of perfect isolation, revealing the above-discussed
spectral structure.  Meanwhile, as the coupling to the bath is increased, the Lorentzian width increases also, erasing structure on ever larger energy scales.
When all structure due to MBL in the local spectral functions is erased by coupling to a bath, the system ceases to show any signs of being MBL. This general picture can be used to consider a wide variety of properties of imperfectly isolated MBL systems, including transport behavior and properties as
a quantum memory.

\begin{figure}
\includegraphics{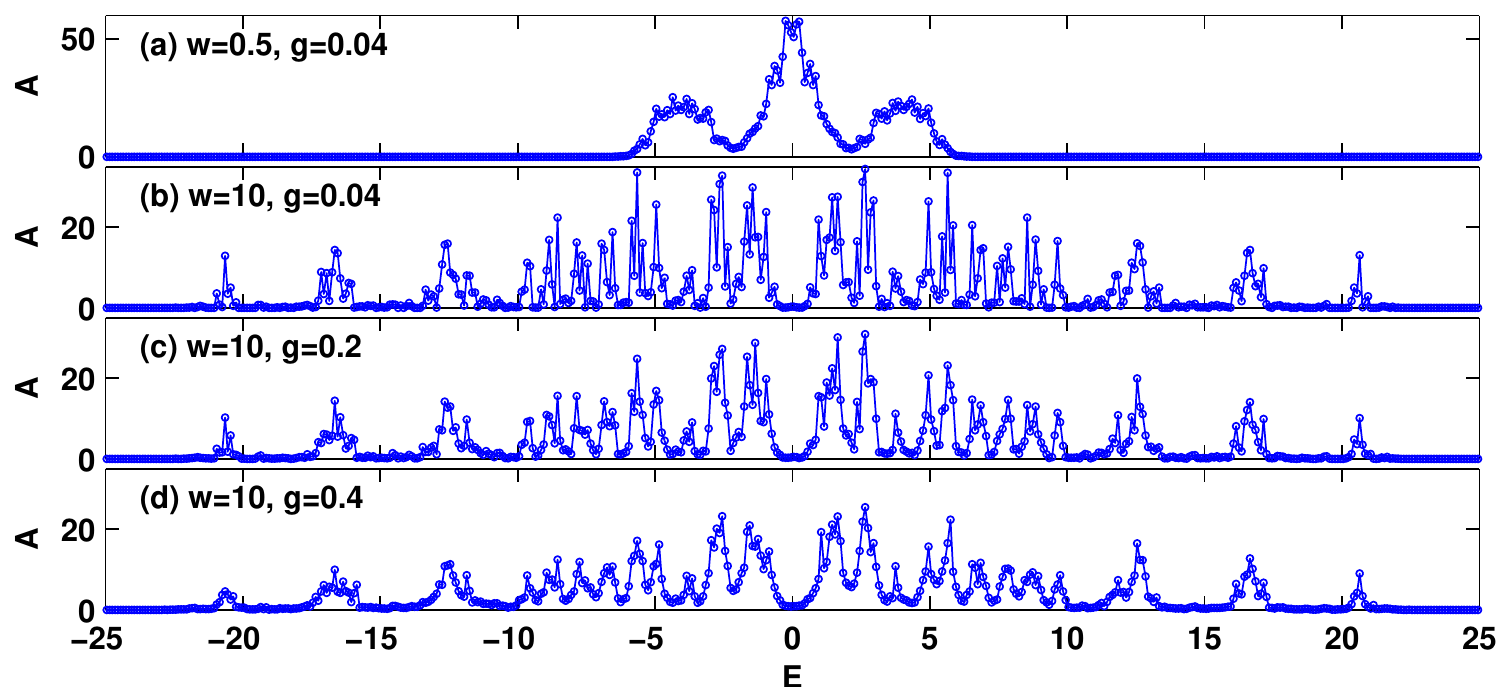}
\caption{\label{fig: spectrum} Figure showing the spectrum of a local spin flip operator $\sigma^x_i$ of a system governed by Hamiltonian (\ref{eq: model2}) coupled to a thermalizing spin chain (a bath) with a coupling $g$.  Figure taken from Ref. \cite{spectralnumerics}.  The spectra are obtained by exact diagonalization, for a system and bath that consist of eight spins each.  The spectra are averaged over spatial position in the system, and also over all many body eigenstates (i.e. the spectra are evaluated in an `infinite temperature' Gibbs mixed state).  Here $w$ controls the disorder strength in the system.  For small $w$ (low disorder), the system is in its thermalizing phase (top panel), and the local spectrum is smooth and continuous.  For large $w$ (lower three panels), the system is in its localized phase if isolated.  For the coupled finite-size system and bath, the eigenstates become effectively thermal above a coupling $g \approx 0.15$.  The second panel shows the local spectrum in the regime where the eigenstates are non-thermal - the local spectrum is highly inhomogenous, and contains a hierarchy of gaps, including the soft gap at zero frequency.  The third and fourth panels show the local spectrum at coupling where the eigenstates of the coupled system and bath are thermal.  We see that although spectral line broadening does smooth out the local spectrum, the local spectrum retains signatures of proximity to localization.}
\end{figure}

These ideas were investigated numerically in \cite{spectralnumerics}, which used exact diagonalization to study the behavior of a spin chain (the `system') with Hamiltonian (\ref{eq: model2}) coupled to a thermalizing spin chain whose eigenstates obey ETH (the `bath'), with a tunable coupling $g$.  For numerical reasons, the study was restricted to considering eight spins or less in the system and eight spins or less in the bath.   When the system and bath were decoupled, the local spectrum of a spin flip operator in the system was found to be very different in the thermal and localized phases (Fig.\ref{fig: spectrum}, first two panels).  On coupling a localized system to the bath, it was found that the eigenstates of the combined system and bath become effectively thermal, with the crossover to thermal eigenstates occurring at a coupling that is exponentially small in the size of the bath.  In contrast, the local spectrum was found to remain `spiky' and inhomogenous even for system-bath couplings where the combined eigenstates were effectively thermal (Fig.\ref{fig: spectrum}, lower two panels). The behavior of the spectrum is illustrated by Fig. \ref{fig: spectrum}, taken from \cite{spectralnumerics}.  Note the existence of a `soft' gap at zero frequency and also the hierarchy of gaps discussed above, and note too how these gaps are gradually filled in as the coupling to the bath is increased.

The discussion in \cite{spectral, spectralnumerics} assumed that the local bandwidth of the bath was larger than the characteristic energy scales in the system.  The behavior in the opposite limit, when the local bandwidth of the bath is smaller than the characteristic energy scales of the system, was investigated in \cite{NarrowBandwidth}.  A qualitatively similar behavior was found, but with a line broadening that scales as a power law function of the bandwidth of the bath.

  \section{Localization protected quantum order}
  \label{lpqo}
  If we look at the exact many-body eigenstates of finite systems, we can ask about their properties as we take the thermodynamic limit.  So far, we have focussed on one question: Are the eigenstates thermal or localized?  But we can also ask about whether these eigenstates exhibit symmetry-breaking or topological order.  For eigenstates that are thermal, the eigenstates will only be ordered when the system is ordered at thermal equilibrium, since the eigenstates are each individually at thermal equilibrium.  But in the many-body localized phase ordering may be present in individual eigenstates that is absent in thermal equilibrium.  This is where the new single-eigenstate statistical mechanics can reveal new phases and phase transitions.
We highlight in particular the possibility of spontaneous symmetry-breaking even below the equilibrium lower critical dimension, and the existence of topological order without a bulk gap.  In both cases an ordered phase that would be destroyed at thermal equilibrium by thermal fluctuations is dynamically protected when those fluctuations are localized and thus static.  Examples of localization protected quantum order have been discussed in \cite{LPQO, Bauer, Pekker, Vosk, Bahri, Chandran}.

\subsection{Example: Ising spin chain}
One of the simpler examples to illustrate localization protected quantum order is a random Ising spin chain, with Hamiltonian
  \begin{equation}
  H = - \sum_{i=1}^L h_i \sigma^x_i - \sum_{i=1}^{L-1}(J_i \sigma^z_i \sigma^z_{i+1} + \lambda_i \sigma^x_i \sigma^x_{i+1}) ~, \label{eq: TFIM}
  \end{equation}
  where the $h_i$, $J_i$ and $\lambda_i$ are nonnegative and are drawn from some probability distributions (say, log-normal).
  This Hamiltonian has an Ising $Z_2$ symmetry which is implemented by the operator $\hat P = \prod_{i=1}^L \sigma^x_i$, that rotates all spins by angle $\pi$ about their $x$-axes.

When the $\{J_i\}$ are all much larger than the $\{h_i,\lambda_i\}$
  the ground state of (\ref{eq: TFIM}) is ferromagnetically ordered.
  More precisely, for any finite $L$, there are two nearly-degenerate Schr\"odinger cat ground states $\approx (|\uparrow \rangle \pm |\downarrow \rangle)/\sqrt{2}$
  that are linear combinations of macroscopically different states that are magnetized up ($|\uparrow\rangle$) and down ($|\downarrow\rangle=\hat P|\uparrow\rangle$).
  The energy difference between these two eigenstates is exponentially small in $L$ and is due to the high-order process where the $\sigma^x$ terms in (\ref{eq: TFIM})
  act to flip the global magnetization between up and down.
  If we start this system in the initial magnetized (symmetry-broken) state $|\uparrow \rangle$ that is a linear combination of these two ground states,
  this state has an exponentially small energy uncertainty, and thus takes a time that is exponentially long in $L$ to evolve away from the symmetry-broken
  $|\uparrow \rangle$ initial state under unitary time evolution.  The divergence of this timescale exponentially with system size is diagnostic of spontaneously
  broken symmetry in the thermodynamic limit.  Alternative diagnostics for spontaneously broken symmetry in a finite-size system include the spin-spin correlations
  $\langle \sigma^z_i \sigma^z_{i+r} \rangle$, which show long-range order in both of the ground states,
  and the system's susceptibility to magnetic fields along the $z$ axis, which diverges exponentially with system size in the ground states.

  While the ground states of the above system can break symmetry, in the absence of static randomness there is no symmetry-breaking at energy densities
  above that of the ground states.
 Such an excited state has a nonzero density of domain walls
 and in the absence of randomness these domain walls are delocalized and dynamically `wander' over the entire chain.
 The presence of such delocalized domain walls means that the equal-time spin correlations
 do not exhibit long-range order.  Equally, a broken-symmetry initial state (e.g. a magnetization pattern) dynamically relaxes through motion of the domain walls.
 The absence of symmetry breaking in the exact eigenstates can be viewed as following from the Landau-Peierls theorem.  The latter states that in one dimension there is no spontaneous symmetry breaking at thermal equilibrium at nonzero temperature, and applies here as long as the system is not many-body localized.

  \subsection{Symmetry-breaking at nonzero energy density with randomness}
  \label{sec: SSB}
  Now let's consider the same system (\ref{eq: TFIM}) with strong $\{J_i\}$ so that the ground states remain ferromagnetic,
  but now with sufficiently strong randomness so that the domain walls
  in its excited states are {\it many-body localized}.  Let $|\phi\rangle$ be a particular
  pattern of $\sigma^z$ magnetization, namely a given pattern of domains and localized domain walls.
  $\hat P |\phi \rangle$ has precisely the same pattern of localized domain walls, but with the local $\sigma^z$ magnetizations all reversed.
  By symmetry, the exact eigenstates in this ordered phase in the limit of large $L$ are Schr\"odinger cat states of the form
  $|n,\pm\rangle \approx (|\phi \rangle_n \pm \hat P |\phi \rangle_n)/\sqrt{2}$, just like the ground states, but each such pair of excited eigenstates
  has own magnetization pattern.
  As a result, the broken-symmetry state $|\phi\rangle_n$ has exponentially small energy uncertainty, and a system prepared in this state
  is metastable, with the magnetization pattern surviving for times exponentially large in system size.  Equivalently, the spin autocorrelation function in any state $\langle \sigma^z_i(0) \sigma^z_i(t)\rangle$ only decays to zero on times exponentially large in system size.
  In this eigenstate ordered phase the system spontaneously breaks the $Z_2$ symmetry in each of its many-body eigenstates.  It is evident from the above argument moreover that the symmetry-breaking is protected by the localization of the domain walls.  The fact that the Landau-Peierls theorem forbids spontaneous symmetry breaking at these energies in one dimension is irrelevant, because the Landau-Peierls theorem assumes thermal equilibrium, so does not apply in this many-body localized phase.

    \begin{figure}
 (a) \includegraphics[width = 0.3\columnwidth]{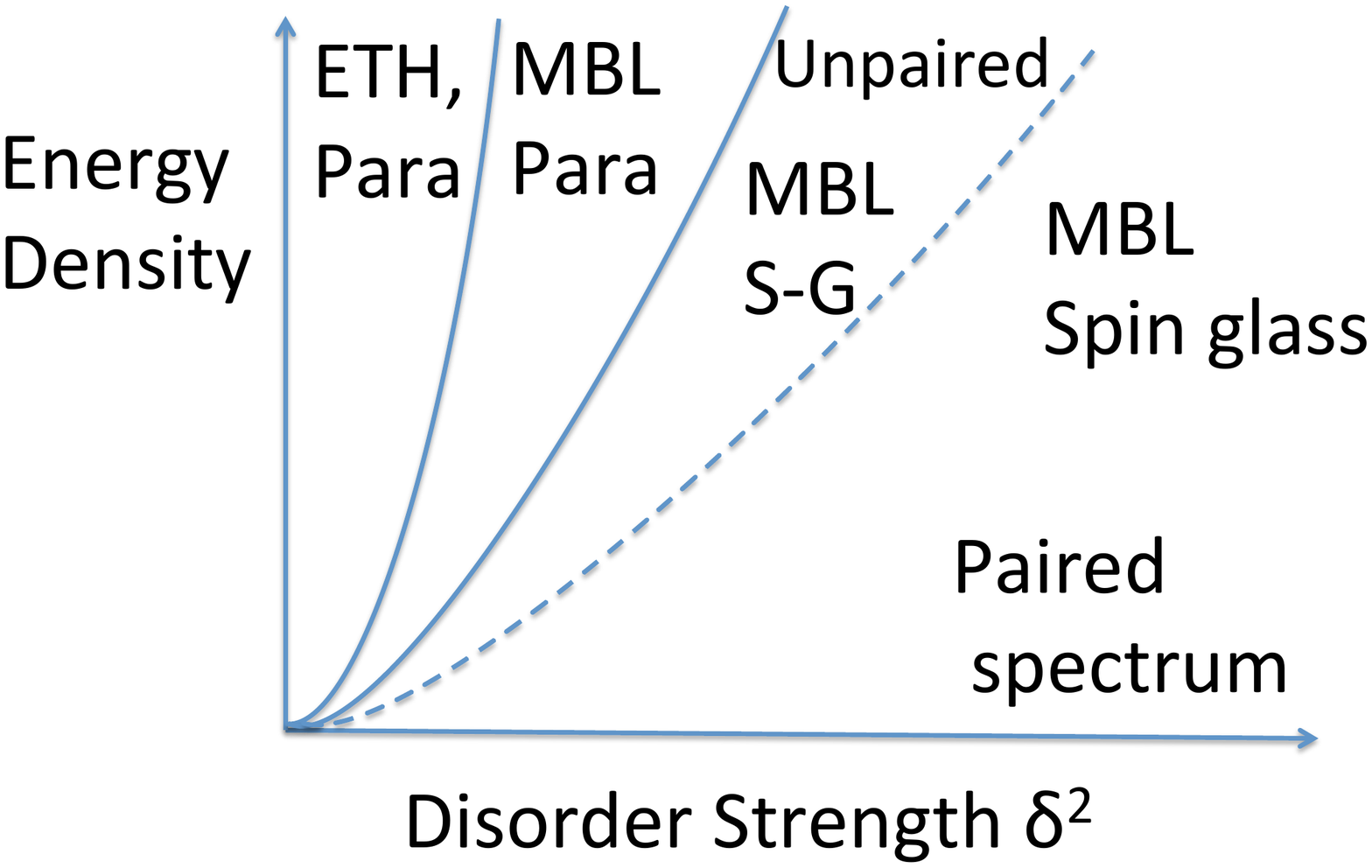}
 (b) \includegraphics[width = 0.25\columnwidth]{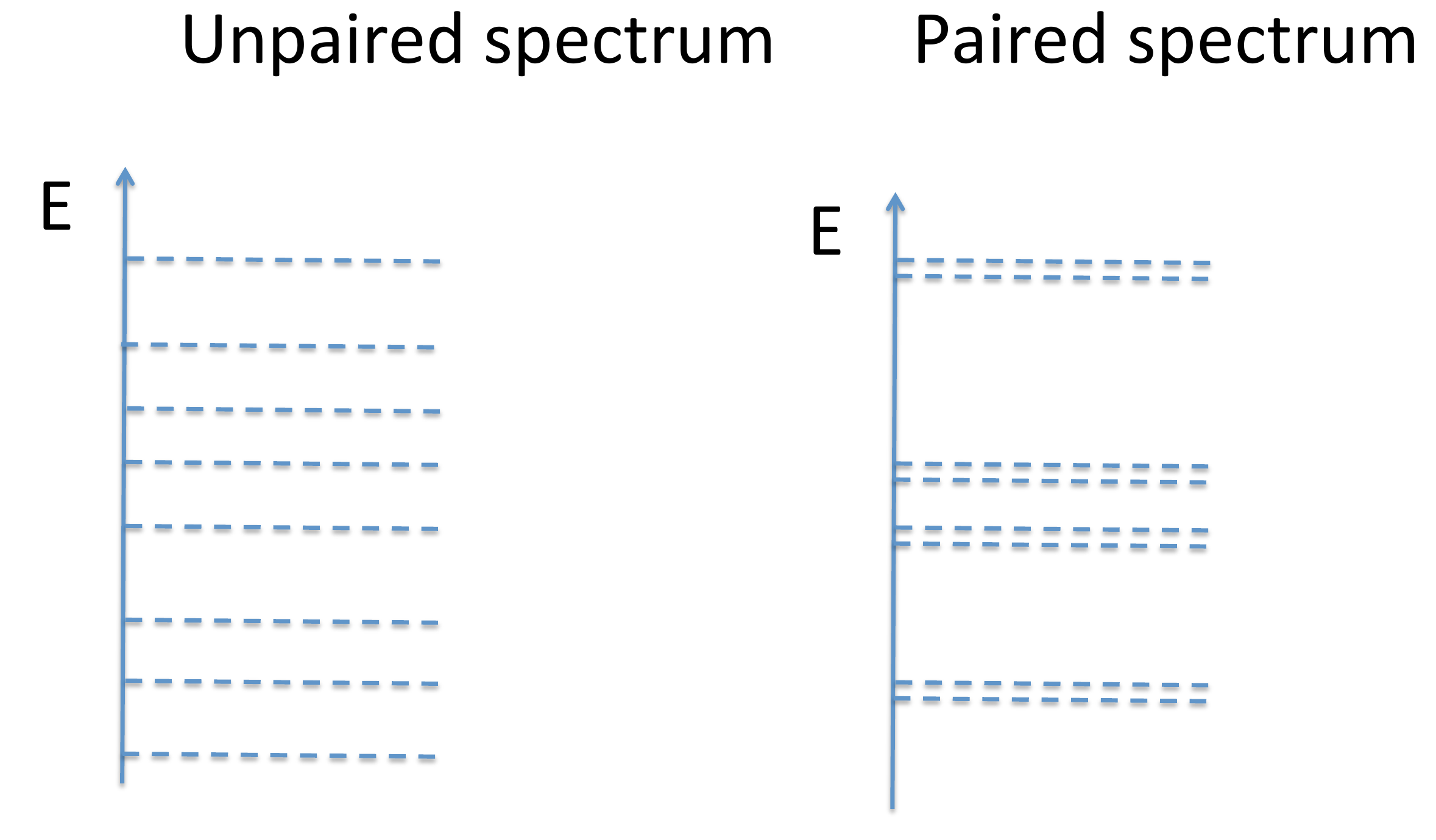}
(c)  \includegraphics[width = 0.33\columnwidth]{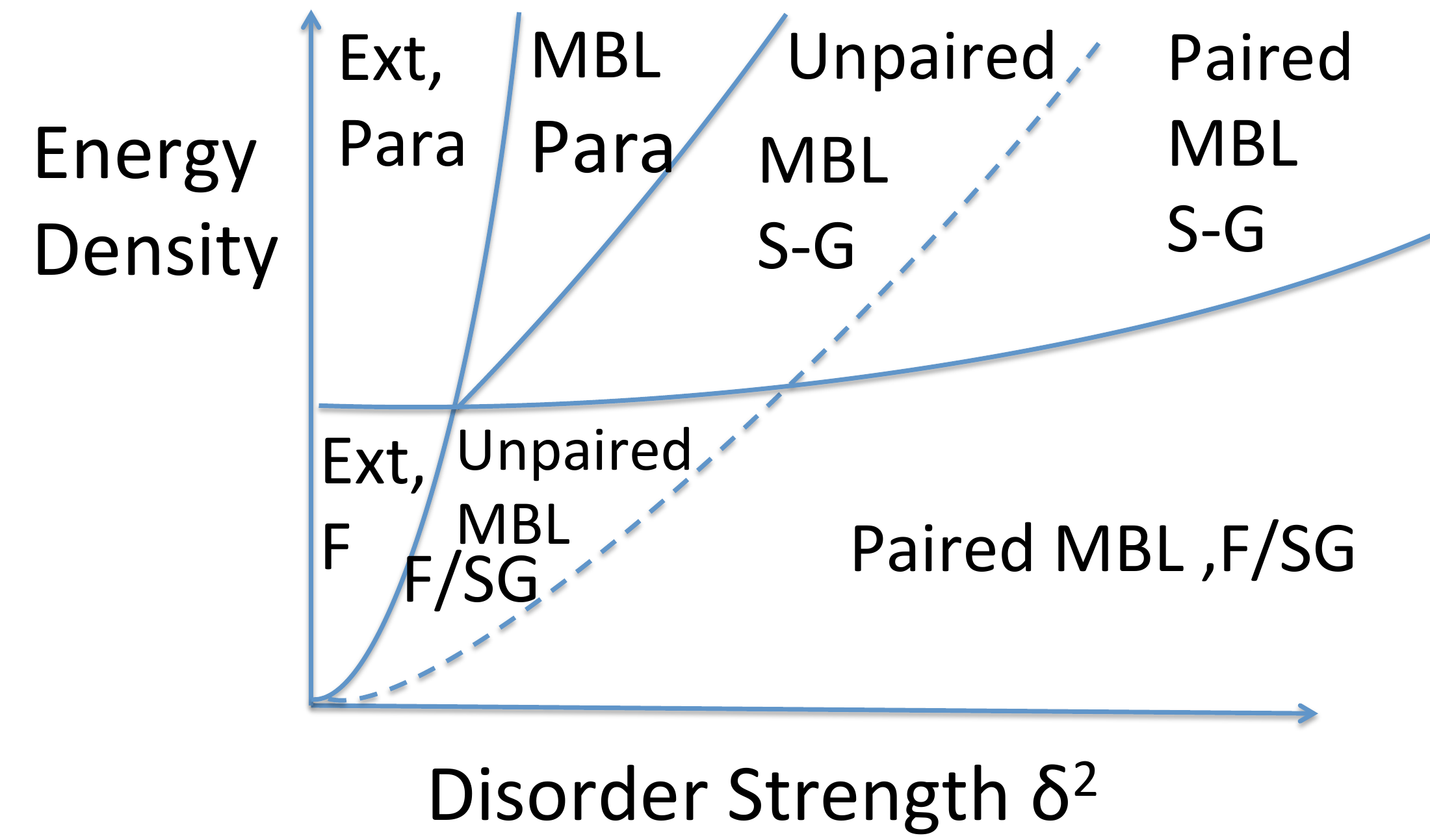}
\caption{\label{fig: SSB} Schematic eigenstate phase diagrams of random quantum Ising models with ferromagnetic ground states.
(a) One dimensional system where there are no thermodynamic phase transitions.  The thermal phase is labeled ETH, while the localized phases are labeled MBL.
Localization stabilizes eigenstate spin glass (S-G) order at energy densities higher than the ground state, even though this is forbidden in thermal equilibrium; see Sec.\ref{sec: SSB}.  The dashed lines in (a) and (c) indicate the spectral transitions between the unpaired phase where the energy splittings between symmetry-related Schr\"odinger cat eigenstates are more than the typical many-body level spacing, and the paired phase where they are less than the typical many-body level spacing. This spectral transition is illustrated in (b).  The eigenstate phase diagram in two dimensions is illustrated in (c).  The principal difference from one dimension (a) is that ferromagnetic order persists for energy densities higher than the ground state.  See Ref. \cite{LPQO} for more detailed discussion of these figures.}
  \end{figure}

To further understand the nature of this broken-symmetry phase, note that the square of the equal-time spin correlation function $\langle \sigma^z_i \sigma^z_{i+r} \rangle^2$, when evaluated in any eigenstate, shows long range order, but $\langle \sigma^z_i \sigma^z_{i+r}\rangle$ changes sign every time we cross one of the localized domain walls. Thus, although these excited eigenstates break the $Z_2$ Ising symmetry, they break it in a `spin glass' manner, with every excited eigenstate being characterized by its own particular $\sigma^z$ magnetization pattern.

Fig. \ref{fig: SSB} illustrates the above discussion.  Note that as we increase the disorder strength at high energy densities, we first transition from
the paramagnetic thermal phase (with only short range spin correlations) to a many-body localized paramagnetic phase.  Here the $x$ components of the local
magnetizations are frozen in nonthermal patterns in each eigenstate, but correlations of the $\{\sigma^z_i\}$ remain short range.  As the disorder strength is
further increased,
there occurs a second transition to a phase with localized domain walls and thus broken $Z_2$ symmetry and long range spin-glass order.  Both of these aforementioned transitions are transitions in the properties of the eigenstates and thus of the system's dynamics, but are not thermodynamic transitions.
This phase diagram has been demonstrated using exact diagonalization data for a one-dimensional model in \cite{MBLMore12}.  The symmetry-breaking transition within the MBL phase has also been studied using the strong-randomness renormalization group \cite{Fisher} in \cite{Pekker,Vosk'}.

Note also the `spectral transition', which occurs within the broken-symmetry MBL phase, and is indicated by the dashed lines in Fig.\ref{fig: SSB}.
As one crosses this spectral transition, the level statistics of the many-body spectrum changes from a Poisson distribution to a `paired' Poisson distribution of symmetry-related doublets composed of pairs of Schr\"odinger cat states.  For further discussion of this spectral transition, see \cite{LPQO}.

\subsection{Localization protected topological order}
\label{sec: to}
We now discuss how localization can also protect topological order at finite energy densities - again even in regimes where such order is forbidden in thermal equilibrium.  We begin by illustrating this in one dimension \cite{LPQO}.  In one dimension, a Jordan-Wigner transformation turns the Ising chain (\ref{eq: TFIM}) into a fermonic model,
\begin{equation}
H =  - \sum_{j=1}^{L-1} i J_j b_j a_{j+1} - \sum_{j=1}^L i h_j a_j b_j + \sum_{j=1}^{L-1} \lambda a_j b_j a_{j+1} b_{j+1} ~, \label{eq: to}
\end{equation}
where the $a_j$ and $b_j$ are Majorana operators. For $\lambda = 0$ and in the absence of disorder $h_i = h$, $J_i = J$, this is simply Kitaev's Majorana chain.
It is known \cite{Fendley} that for $\lambda = 0$ without disorder in the regime $J>h$ there is a fermion mode which is {\it bilocalized}
as two Majorana modes at the two ends of the
chain, and which has energy $\sim\exp(-L\ln J/h)$.  Thus, all the eigenstates of the clean Hamiltonian, for $\lambda = 0$ and $J>h$, come in pairs,
which differ only by the occupation of this fermion edge mode.  Colloquially, `all the eigenstates have Majorana edge modes'.

In the ground state, these Majorana edge modes survive nonzero ($\lambda\neq 0$) interactions, because they are protected by the bulk gap.  However, in the excited states, there is no bulk gap.  Indeed, in the absence of static randomness, the excited states consist of delocalized bulk excitations, which for $\lambda\neq 0$ interact with the edge Majoranas, and thus mediate interactions between the two Majorana modes at the two ends of the system.  As a result, the edge Majorana modes are not robust to non-zero $\lambda$ in the excited states.

In contrast, if the Dirac quasiparticles in the bulk are localized on disorder, then they cannot mediate an interaction between the two ends of the wire. Thus, we expect that in the many-body localized phase, the excited states of the Hamiltonian (\ref{eq: to}) can also carry edge Majoranas for $\lambda\neq 0$, and thus exhibit topological order at nonzero energy density above the ground state.  This is simply the dual of the statement that the Ising spin chain has spin-glass ordered Schr\"odinger cat eigenstates.

A richer manifestation of topological order is found in two dimensions \cite{LPQO, Bauer}. In two dimensions the quantum Ising model has a dual description in terms of a $Z_2$ lattice gauge theory \cite{Fradkin}.  The paramagnetic phase of the Ising model is dual to a deconfined phase of the $Z_2$ lattice gauge theory, with spin flip excitations being dual to `visons' (topological charges).  The localization of spin flips (topological charges) on quenched disorder allows a `spin glass' version of topological order to persist even in highly-excited states of the lattice gauge theory.  This topological order can be diagnosed by now familiar measures, such as Wilson loops, the Fredenhagen-Marcu order parameter, eigenstate degeneracy on a torus, and topological entanglement entropy \cite{LPQO}.

In addition to symmetry-breaking and topological order, in certain other systems there can exist `symmetry protected topological orders', which have recently received much attention in the theory literature \cite{Senthil}.  Symmetry protected topological order in ground states of such nonrandom systems is protected by a bulk gap (as well as by certain symmetries).  It has been pointed out that just as localization can protect topological order in highly-excited states, localization can also protect symmetry protected topological order.  For a fuller discussion of these issues, see \cite{Bahri, Chandran}.

Fermionic systems with topological bands (such as the integer quantum Hall system) present an additional complication, because of the well known topological obstruction to constructing fully localized Wannier orbitals. In such systems, in the non-interacting limit, the single-particle localization length $\xi_{SP}$ diverges near one or more critical single-particle energies $\{E_c\}$ with a critical exponent $\nu$, as $\xi_{SP} \sim |E-E_c|^{-\nu}$. The stability of localization in such systems was recently investigated in \cite{RNACP}. It was established that arbitrarily weak short range interactions trigger delocalization in partially filled bands at non-zero energy density if $\nu > 1/d$. Since general arguments \cite{Harris, Chayes} constrain $\nu \ge 2/d$, it appears that many-body localization can not occur in such systems. For a further discussion of these issues, see \cite{RNACP}.

\section{Some open questions}

One could argue that the most important open question in many-body localization is:  What are the experimental systems and techniques with which this phenomenon will be investigated in the laboratory?  Some ideas about this are discussed in e.g. \cite{Serbynetal,agkl,Mirlin,BAA,MBLMore10,expts1,expts2,expts3}.  But this is a review of theoretical issues, so we now instead consider some of the interesting open theoretical questions.

\subsection{Localization in translationally-invariant systems}
\label{annealed}
One interesting class of open questions revolves around the issue of whether
many-body localization or something like it can occur also in systems with translationally-invariant Hamiltonians.  Can randomness in the initial {\it state} of the system be enough to `localize itself', even though the Hamiltonian is translationally-invariant?  If so, could quantum localization thus be relevant for, say, structural glasses, where certain degrees of freedom within a glassy material are in some sense `self-localized' and do not thermalize?

There are a variety of related types of models that have been considered here.  One comes from considering the motion of $^3$He `impurities' in solid $^4$He \cite{Kagan}.  If one assumes power-law long-range interactions between the impurities, then the hop of a single impurity atom to a neighboring lattice site changes the system's energy by some amount that depends on the positions of all other nearby impurities.  Random initial locations of all the impurities are thus argued \cite{Kagan} to localize all the impurity atoms if the hopping is weak enough.  Another variation on this idea uses a generalized Bose-Hubbard model \cite{Huveneers}, with the randomness in the initial state being instead the large and random occupation numbers on each site.  In this case the authors proved an upper bound on the transport, demonstrating that the transport is at most nonperturbative in the hopping; they call this `asymptotic localization' \cite{Huveneers}.

Another class of models \cite{gf, Schiulaz} has two species of particles, one light and one heavy.  Given a random initial state of the positions of the heavy particles, one can then consider a `Born-Oppenheimer' approximation to the states of the light particles, and if the heavy particles produce enough disorder or enough constraints the light particles are localized by the (now stationary) heavy particles.  For a lattice model, one can then allow hopping of the heavy particles, and argue that the transport is again at most nonperturbative in this hopping \cite{Schiulaz}.

An open question here is whether such translationally-invariant systems can have true many-body localization, with strictly zero DC transport at some nonzero hopping, or are they always `only' asymptotically localized, with some nonperturbative effect producing nonzero transport at any nonzero hopping?  One possible source of nonperturbative transport is rare regions where the state is locally much less random and thus has mobile excitations \cite{HdR, mbglass}.  If these rare regions themselves are mobile, they could conceivably thermalize the system.  But even if such systems can only have asymptotic localization, they will still have a large regime of weak hopping where systems of finite size will in many ways appear to be localized.

Very recently, a new connection between many-body localization and structural glass has been drawn by \cite{Garrahan}.  This work studies a particular stochastic classical dynamics which is known to have a glass transition, and maps the classical master equation governing this dynamics to a quantum Hamiltonian at a Rokhsar-Kivelson (RK) point \cite{RK}.  A numerical investigation reveals that the many-body eigenstates of the quantum Hamiltonian violate the ETH both at the RK point, and apparently also for small perturbations away from it.  The connection to other works on many-body localization in translationally-invariant systems \cite{gf, Schiulaz, Huveneers, HdR, mbglass} is not presently clear.

\subsection{The many-body localization phase transition}
\label{sec: phase transition}

Another set of open questions is about the nature (the universality class) of the many-body localization phase transition between the thermal and localized phases as the randomness is increased.  The first question is whether there is only one phase transition, or could there possibly be some sort of intermediate phase that is neither fully localized nor fully thermal \cite{Grover}?
Studies of this transition so far have mostly been numerical studies based on exact diagonalization of relatively short one-dimensional systems \cite{Oganesyan,pal,MBLMore12}.
Very little is known about this transition, so how to do a proper finite-size scaling analysis of these numerical data remains unclear.  One of these studies did investigate dynamic scaling and probability distributions near the transition, suggesting dynamic critical exponent $z\rightarrow\infty$ and the possibility that the transition is governed by an infinite-randomness fixed point \cite{pal}.  This suggests that the transition might be treated with some version of a strong-randomness renormalization group.

Recently, the strong sub-additivity of entanglement entropy has been used to establish \cite{Grover} that if there is a direct continuous phase transition from a thermal phase obeying the ETH to an MBL phase, then the ETH remains true {\it at} the phase transition.  This result constrains the possible theories of the many-body localization transition.

A different approach was taken in \cite{NarrowBandwidth}, which examined the local spectral functions in the thermal phase in the vicinity of the localization transition.  The assumption in \cite{NarrowBandwidth} was that since the local spectra are discrete in the localized phase and continuous in the thermal phase, the local spectra in a thermal phase that is close to localization should become exceedingly inhomogenous, and may be characterized by a spectral line width which goes to zero as the localized phase is approached.  The spectral line width, which is proportional to observables such as the DC conductivity, may then be viewed as an order parameter for the transition.  The scaling of the line width with control parameters such as disorder strength, interaction strength and temperature was worked out in a self-consistent mean-field approximation.  However, any such mean-field theory is expected to fail in the immediate vicinity of the critical point, so a description of the critical physics remains an open problem.  Progress on understanding more about this phase transition remains an outstanding challenge.

\label{transition}

\section{Conclusion}
In this review we have surveyed some of the present theoretical understanding of quantum thermalization and many-body Anderson localization.  We emphasize that although significant progress has been made in understanding these phenomena in recent years, there still remain many open issues, and much of the present `understanding' is really in the form of conjectures or hypotheses that need to be either verified or replaced with more accurate ideas.  Thus
this is a field that is still in its infancy.  This particularly applies to possible experimental investigations of many-body localization,
where the field is still so young that it seemed to us too early to attempt a review of the possibilities.
We look forward to new researchers joining in to work on these topics and thus advancing the future progress of this field.

\section{Acknowledgements}

We thank Ehud Altman, Boris Altshuler, Ravin Bhatt, J\"urg Fr\"ohlich, Sarang Gopalakishnan, John Imbrie, Sonika Johri, Vedika Khemani, Hyungwon Kim, Joel Lebowitz, Marcus M\"uller, Vadim Oganesyan, Arijeet Pal, Andrew Potter, Gil Refael, Shivaji Sondhi, Tom Spencer and Ronen Vosk for collaborations and/or discussions.


\begin{thebibliography}{99}
\bibitem[Sakurai (1983)]{sakurai}
J. J. Sakurai, {\it Modern Quantum Mechanics}, San Fu Tauan (Benjamin/Cummings, Menlo Park, CA 1985).
\bibitem{npi}
See, e.g., Nature Physics Insight {\bf 8}, 263-299 (2012).
\bibitem[Kardar (2007)]{Kardar}
M. Kardar, {\it Statistical Physics of Particles}, 1st ed., Cambridge University Press (2007).
\bibitem{bloch}
E.g., I. Bloch, J. Dalibard and S. Nascimbéne, Nat. Phys. {\bf 8}, 267 (2012).
\bibitem[PolkovnikovRMP (2011)]{PolkovnikovRMP}
A. Polkovnikov, K. Sengupta, A. Silva, M. Vengalatorre, Rev. Mod. Phys. {\bf 82}, 031130 (2010).
\bibitem[Anderson (1958)]{Anderson}
P. W. Anderson,
Phys. Rev. {\bf 109}, 1492 (1958).
\bibitem[LPQO (2013)]{LPQO}
D. A. Huse, R. Nandkishore, V. Oganesyan, A. Pal and S. L. Sondhi, Phys. Rev. B {\bf 88}, 014206 (2013).
\bibitem[Fleishman (2366)]{Fleishman}
L. Fleishman and P. W. Anderson, Phys. Rev. B {\bf 21}, 2366 (1980).
\bibitem[AKGL (1997)]{agkl}
B. L. Altshuler, Y. Gefen, A. Kamenev and L. S. Levitov, Phys. Rev. Lett. {\bf 78}, 2803 (1997).
\bibitem[Mirlin (2006)]{Mirlin}
I. V. Gornyi, A. D. Mirlin and D. G. Polyakov, Phys. Rev. Lett. {\bf 95}, 206603 (2005).
\bibitem[BAA (2006)]{BAA}
D. M. Basko, I. L. Aleiner and B. L. Altshuler, Annals of Physics {\bf 321}, 1126 (2006).
\bibitem[Oganesyan (2008)]{Oganesyan}
V. Oganesyan and D. A. Huse,
Phys. Rev. B {\bf 75}, 155111 (2007).
\bibitem[Pal (2010)]{pal}
A. Pal and D. A. Huse,
Phys. Rev. B {\bf 82}, 174411 (2010).
\bibitem{IS}
J. Z. Imbrie, arXiv:1403.7837.
\bibitem[Bauer (2013)]{Bauer}
B. Bauer and C. Nayak, J. Stat. Mech. P09005 (2013).
\bibitem[Vosk' (2013)]{Vosk'}
R. Vosk and E. Altman, Phys. Rev. Lett. {\bf 110}, 067204 (2013).
\bibitem[Pekker (2013)]{Pekker}
D. Pekker, G. Refael, E. Altman, E. Demler and V. Oganesyan, Phys. Rev. X {\bf 4}, 011052 (2014).
\bibitem[Vosk (2013)]{Vosk}
R. Vosk and E. Altman, arXiv:1307.3256.
\bibitem[Bahri (2013)]{Bahri}
Y. Bahri, R. Vosk, E. Altman and A. Vishwanath, arXiv:1307.4192.
\bibitem[Chandran (2013)]{Chandran}
A. Chandran, V. Khemani, C. R. Laumann and S. L. Sondhi, Phys. Rev. B {\bf 89}, 144201 (2014).
\bibitem[Abanin (2013)]{Abanin1}
M. Serbyn, Z. Papic and D. A. Abanin, Phys. Rev. Lett. {\bf 110}, 260601 (2013).
\bibitem[Lbits (2013)]{Lbits}
D. A. Huse and V. Oganesyan, arXiv:1305.4915.
\bibitem[Abanin (2013)]{Abanin2}
M. Serbyn, Z. Papic and D. A. Abanin, Phys. Rev. Lett. {\bf 111}, 127201 (2013).
\bibitem[Lbits2 (2013)]{Lbits2}
D. A. Huse, R. Nandkishore and V. Oganesyan, in preparation.
\bibitem[MBLMore (2008)]{MBLMore1}
M. Znidaric, T. Prosen and P. Prelovsek, Phys. Rev. B {\bf 77}, 064426 (2008).
\bibitem[MBLMore (2010)]{MBLMore2}
C. Monthus and T. Garel, Phys. Rev. B {\bf 81}, 134202 (2010).
\bibitem[MBLMore (2010)]{MBLMore3}
T. C. Berkelbach and D. R. Reichman, Phys. Rev. B {\bf 81}, 224429 (2010).
\bibitem[MBLMore (2011)]{MBLMore4}
E. Canovi, D. Rossini, R. Fazio, G. E. Santoro and A. Silva, Phys. Rev. B {\bf 83}, 094431 (2011); New J. Phys. {\bf 14}, 095020 (2012).
\bibitem[MBLMore (2012)]{MBLMore5}
M. V. Feigel'man, L. B. Ioffe and M. Mezard, Phys. Rev. B {\bf 82}, 184534 (2010).
\bibitem[MBLMore (2013)]{MBLMore6}
A. De Luca and A. Scardicchio, Europhys. Lett. {\bf 101}, 37003 (2013).
\bibitem[MBLMore (2013)]{MBLMore7}
B. Swingle, arXiv:1307.0507.
\bibitem[MBLMore (2013)]{MBLMore8}
R. Sims and G. Stolz, arXiv:1312.0577.
\bibitem[MBLMore (2013)]{MBLMore9}
Y. Bar Lev and D. R. Reichman, arXiv:1402.0502.
\bibitem[MBLMore (2010)]{MBLMore10}
I. L. Aleiner, B. L. Altshuler and G. V. Shlyapnikov, Nat. Phys. {\bf 6}, 900 (2010).
\bibitem[MBLMore (2014)]{MBLMore11}
V. P. Michal, B. L. Altshuler and G. V. Shlyapnikov, arXiv:1402.4796.
\bibitem[MBLMore (2014)]{MBLMore12}
J. A. Kj\"all, J. H. Bardarson and F. Pollman, arXiv:1403.1568.
\bibitem[MBLMore (2014)]{MBLMore13}
E. Khatami, M. Rigol, A. Relano, A. M. Garcia-Garcia, Phys. Rev. E {\bf 85}, 050102(R) (2012)
\bibitem[Floquet (2013)]{Floquet}
L. D'Alessio and A. Polkovnikov, Annals of Physics {\bf 333}, 19 (2013).
\bibitem[Floquet2 (2014)]{Floquet2}
P. Ponte, A. Chandran, Z. Papic and D. A. Abanin, arXiv: 1403.6480
\bibitem[GGE0 (2007)]{GGE0}
M. Rigol, V. Dunjko, V. Yurovsky and M. Olshanii, Phys. Rev. Lett. {\bf 98}, 050405 (2007).
\bibitem[GGE (2012)]{GGE1}
A. C. Cassidy, C. W. Clark and M. Rigol, Phys. Rev. Lett. {\bf 106}, 140405 (2011).
\bibitem[GGE (2013)]{GGE2}
J. S. Caux and F. H. L. Essler, Phys. Rev. Lett. {\bf 110}, 257203 (2013).
\bibitem[lych (2013)]{lych}
O. Lychkovskiy, Phys. Rev. A {\bf 87}, 022112 (2013).
\bibitem[Deutsch (1991)]{Deutsch}
J. M. Deutsch, Phys. Rev. A {\bf 43}, 2146 (1991).
\bibitem[Srednicki (1994)]{Srednicki}
M. Srednicki, Phys. Rev. E {\bf 50}, 888 (1994).
\bibitem{tasaki}
H. Tasaki, Phys. Rev. Lett. {\bf 80}, 1373 (1998).
\bibitem[Rigol (2008)]{Rigol}
M. Rigol, V. Dunjko and M. Olshanii, Nature {\bf 452}, 854 (2008).
\bibitem[ETH (2009)]{ETH1}
M. Rigol, Phys. Rev. Lett. {\bf 103}, 100403 (2009).
\bibitem[ETH (2010)]{ETH2}
M. Rigol and L. F. Santos, Phys. Rev. A {\bf 82}, 011604(R) (2010).
\bibitem[ETH (2011)]{ETH3}
T. N. Ikeda, Y. Watanabe and M. Ueda, Phys. Rev. E {\bf 84}, 021130 (2011); arXiv:1202.1965.
\bibitem[ETH (2012)]{ETH4}
M. Rigol and M. Srednicki, Phys. Rev. Lett. {\bf 108}, 110601 (2012).
\bibitem[ETH (2012)]{ETH5}
S. Dubey, L. Silvestri, J. Finn, S. Vinjanampathy and K. Jacobs, Phys. Rev. E {\bf 85}, 011141 (2012).
\bibitem[ETH (2013)]{ETH6}
R. Steinigeweg, J. Herbych and P. Prelovsek, Phys. Rev. E {\bf 87}, 012118 (2013).
\bibitem[ETH (2013)]{ETH7}
A. De Luca, arXiv:1302.0992.
\bibitem{ETH7.5}
W. Beugeling, R. Moessner and M. Haque, Phys. Rev. E {\bf 89}, 042112 (2014).
\bibitem[ETH (2013)]{ETH8}
R. Steinigeweg, H. Niemeyer, C. Gogolin and J. Gemmer, Phys. Rev. Lett. {\bf 112}, 130403 (2014).
\bibitem[ETH (2013)]{ETH9}
S. Khlebnikov and M. Kruczenski, arXiv:1312.4612.
\bibitem[ETH (2013)]{ETH10}
M. P. Mueller, E. Adlam, L. Masanes, and N. Wiebe, arXiv:1312.7420.
\bibitem[localizationRMP (2008)]{localizationRMP}
F. Evers and A. D. Mirlin, Rev. Mod. Phys. {\bf 80}, 1355 (2008).
\bibitem{Iyer}
S. Iyer, V. Oganesyan, G. Refael and D. A. Huse, Phys. Rev. B {\bf 87}, 134202 (2013).
\bibitem[Serbynetal (2014)]{Serbynetal}
M. Serbyn, M. Knap, S. Gopalakrishnan, Z. Papic, N. Y. Yao, C. R. Laumann, D. A. Abanin, M. D. Lukin and E. A. Demler, arXiv:1403.0693.
\bibitem[Bardarson (2012)]{Bardarson}
J. H. Bardarson, F. Pollman and J. E. Moore, Phys. Rev. Lett. {\bf 109}, 017202 (2012).
\bibitem{kim}
H. Kim and D. A. Huse, Phys. Rev. Lett. {\bf 111}, 127205 (2013).
\bibitem{vah}
It appears that the entanglement spread can instead be a sub-ballistic power law of time in one-dimensional thermalizing systems, if `weak links' due to rare, almost insulating regions are common enough to render the entanglement spread sub-ballistic: R. Vosk, E. Altman and D. A. Huse, work in progress.
\bibitem{Lieb}
E.H. Lieb and D. Robinson, {\it Commun. Math. Phys.} {\bf 28}, 251-257 (1972)
\bibitem{bm}
A. J. Bray and M. A. Moore, Phys. Rev. Lett. {\bf 58}, 57 (1987).
\bibitem[Spectral (2014)]{spectral}
R. Nandkishore, S. Gopalakrishnan and D. A. Huse, arXiv:1402.5971.
\bibitem[SpectralNumerics (2014)]{spectralnumerics}
S. Johri, R. Nandkishore and R. N. Bhatt, arXiv:1405.5515.
\bibitem[NarrowBandwidth (2014)]{NarrowBandwidth}
S. Gopalakrishnan and R. Nandkishore, arXiv:1405.1036.
\bibitem[Fisher (2012)]{Fisher}
D. S. Fisher, Phys. Rev. B {\bf 51}, 6411 (1995).
\bibitem[Fendley (2012)]{Fendley}
P. Fendley, J. Stat. Mech. P11020 (2012).
\bibitem[Fradkin (1979)]{Fradkin}
E. Fradkin and S. H. Shenker, Phys. Rev. D {\bf 19}, 3682 (1979).
\bibitem[Sentil (2012)]{Senthil}
A. Vishwanath and T. Senthil, Phys. Rev. X {\bf 3}, 011016 (2013), and references contained therein.
\bibitem[RNACP (2014)]{RNACP}
R. Nandkishore and A. C. Potter, arXiv:1406.0847.
\bibitem[Harris (1980)]{Harris}
A. B. Harris, J. Phys. C {\bf 7}, 1671 (1974).
\bibitem[Chayes (1986)]{Chayes}
J. T. Chayes, L. Chayes, D. S. Fisher and T. Spencer, Phys. Rev. Lett. {\bf 57}, 2999 (1986).
\bibitem[expts (2013)]{expts1}
D. M. Basko, I. L. Aleiner and B. L. Altshuler, Phys. Rev. B 76, 052203 (2007).
\bibitem[expts (2013)]{expts2}
M. P. Kwasigroch and N. Cooper, arXiv:1311.5393.
\bibitem[exots (2013)]{expts3}
N. Y. Yao, C. R. Laumann, S. Gopalakrishnan, M. Knap, M. Mueller, E. A. Demler, M. D. Lukin, arXiv:1311.7151.
\bibitem[Kagan (1984)]{Kagan}
Yu. Kagan and L. A. Maksimov, Zh. Eksp. Teor. Fiz. {\bf 87}, 348-365 (1984); Sov. Phys. JETP {\bf 60 (1)}, 201-210 (1984).
\bibitem[Huveneers (2014)]{Huveneers}
F. Huveneers and W. De Roeck, arXiv:1308.6263.
\bibitem{gf}
T. Grover and M. P. A. Fisher, arXiv:1307.2288.
\bibitem[Schiulaz (2013)]{Schiulaz}
M. Schiulaz, M. Muller, arXiv:1309.1082.
\bibitem[HdR (2014)]{HdR}
F. Huveneers and W. De Roeck, arXiv:1405.3279.
\bibitem[mbglass (2014)]{mbglass}
R. Nandkishore, D. A. Huse and S. L. Sondhi, in preparation.
\bibitem[Garrahan (2014)]{Garrahan}
J. M. Hickey, S. Genway and J. P. Garrahan, arXiv:1405.5780.
\bibitem[RK (2005)]{RK}
C. Castelnovo, C. Chamon, C. Mudry and P. Pujol, Annals of Physics, {\bf 318}, 2, 306-344 (2005), and references contained therein.
\bibitem[Grover (2014)]{Grover}
T. Grover, arXiv:1405.1053.
\end{thebibliography}
\end{document}